\documentclass[iop]{emulateapj}
\usepackage{graphicx,subfigure,amsmath, amsfonts, amssymb,aas_macros,footnote,color,epstopdf}
\usepackage[bookmarks=false]{hyperref}
\hypersetup{
  colorlinks= true, 
  urlcolor  = black, 
  linkcolor = red, 
  citecolor = blue 
} 
\usepackage{multirow}
\newcommand{\msun}{M$_{\odot}$}

\newcommand{\atlas}{ATLAS$^{\rm 3D}$}

\shorttitle{VLBA Imaging of NGC~1266}
\shortauthors{K. Nyland et al.}

\begin{document}

\title{Detection of a High Brightness Temperature Radio Core in the AGN-Driven Molecular Outflow Candidate NGC~1266}

\author{Kristina Nyland\altaffilmark{1},\email{knyland@nmt.edu} Katherine Alatalo\altaffilmark{2,3}, J.~M. Wrobel\altaffilmark{4}, Lisa M. Young\altaffilmark{1}, Raffaella Morganti\altaffilmark{5, 6}, Timothy A. Davis\altaffilmark{7}, P.~T. de Zeeuw\altaffilmark{7,8}, Susana Deustua\altaffilmark{9}, Martin Bureau\altaffilmark{10}}
\altaffiltext{1}{Physics Department, New Mexico Tech, Socorro, NM 87801, USA; knyland@nmt.edu}
\altaffiltext{2}{Department of Astronomy, Hearst Field Annex, University of California - Berkeley, California 94720, USA}
\altaffiltext{3}{Infrared Processing and Analysis Center, California Institute of Technology, Pasadena, California 91125, USA}
\altaffiltext{4}{National Radio Astronomy Observatory, Socorro, NM 87801, USA}
\altaffiltext{5}{Netherlands Institute for Radio Astronomy, Postbus 2, 7990 AA, Dwingeloo, The Netherlands}
\altaffiltext{6}{Kapteyn Astronomical Institute, University of Groningen, Postbus 800, 9700 AV Groningen, The Netherlands}
\altaffiltext{7}{European Southern Observatory, Karl-Schwarzschild-Str.\ 2, D-85748 Garching, Germany}
\altaffiltext{8}{Sterrewacht Leiden, Leiden University, Postbus 9513, 2300 RA Leiden, The Netherlands}
\altaffiltext{9}{Space Telescope Science Institute, 3700 San Martin Drive, Baltimore, MD 21218, USA}
\altaffiltext{10}{Sub-department of Astrophysics, Department of Physics, University of Oxford, Denys Wilkinson Building, Keble Road, Oxford OX1 3RH, UK}

\begin{abstract}
We present new high spatial resolution Karl G. Jansky Very Large Array (VLA) H{\tt I} absorption and Very Long Baseline Array (VLBA) continuum observations of the Active Galactic Nucleus (AGN)-driven molecular outflow candidate NGC 1266.  Although other well-known systems with molecular outflows may be driven by star formation in a central molecular disk, the molecular mass outflow rate reported in Alatalo et al.\ (2011) in NGC 1266 of 13 M$_{\odot}$ year$^{-1}$ exceeds star formation rate estimates from a variety of tracers.  This suggests that an additional energy source, such as an AGN, may play a significant role in powering the outflow.  Our high spatial resolution H{\tt I} absorption data reveal compact absorption against the radio continuum core co-located with the putative AGN, and the presence of a blueshifted spectral component re-affirms that gas is indeed flowing out of the system.  Our VLBA observations at 1.65~GHz reveal one continuum source within the densest portion of the molecular gas, with a diameter d $<$ 8 mas (1.2~pc), a radio power $P_{\mathrm{rad}}$ = 1.48 $\times$ 10$^{20}$ W~Hz$^{-1}$, and a brightness temperature $T_{\mathrm{b}} >$ 1.5 $\times$ 10$^7$ K that is most consistent with an AGN origin.  The radio continuum energetics implied by the compact VLBA source, as well as archival VLA continuum observations at lower spatial resolution, further support the possibility that the AGN in NGC~1266 could be driving the molecular outflow.  These findings suggest that even low-level AGNs may be able to launch massive outflows in their host galaxies.  
\end{abstract}



\section{Introduction}
Simulations in recent years \citep{springel+05, Hopkins+05, croton+06, debuhr+12, dubois+13} have suggested that active galactic nuclei (AGNs) may be able to quench star formation (SF) and play a significant role in the buildup of ``red and dead" galaxies.  At redshifts of a few, this process is expected to be predominantly carried out through radiative feedback in the so-called ``quasar mode" \citep{ciotti+07}, while mechanical feedback in the ``radio mode" is believed to dominate less energetic systems in the local universe \citep{ciotti+10}.  In both feedback modes, the subsequent large-scale expulsions of molecular gas are expected to directly stifle SF by clearing out the raw materials necessary for new generations of stars.  

During this relatively short stage of galaxy evolution, galactic nuclei are expected to be obscured by thick column densities of molecular gas, making it difficult to extract direct, observational evidence implicating the AGN as the primary driver of the outflow (e.g., \citealt{fabian+12}).  Known candidates for AGN-driven SF quenching have thus remained controversial and primarily consist of Ultraluminous Infrared Galaxies (ULIRGS; e.g., \citealt{feruglio+10, sturm+11}) as well as more distant quasars (e.g., \citealt{nesvadba+08, cano-diaz+12, greene+12, maiolino+12}).  It has been suggested that the less powerful AGNs that populate the nearby universe may also be capable of driving molecular outflows and shutting down late-time SF in some local galaxies \citep{schawinski+09, chung+11}, and the number of such candidates has been increasing \citep{das+05, matsushita+07, alatalo+11, dasyra+12, hota+12, aalto+12b, morganti+13, sakamoto+13}.  Of these intriguing local candidates for AGN-driven SF quenching, NGC~1266 \citep{alatalo+11, alatalo+13, davis+12} provides a rare example of a host galaxy that is both non-starbursting and shows no evidence of a recent major merger or interaction (although see Aalto et al.\ (2012b) and Sakamoto et al.\ (2013) for other possible examples).

 \begin{figure*}[t!]
\centering
\includegraphics[width=7in]{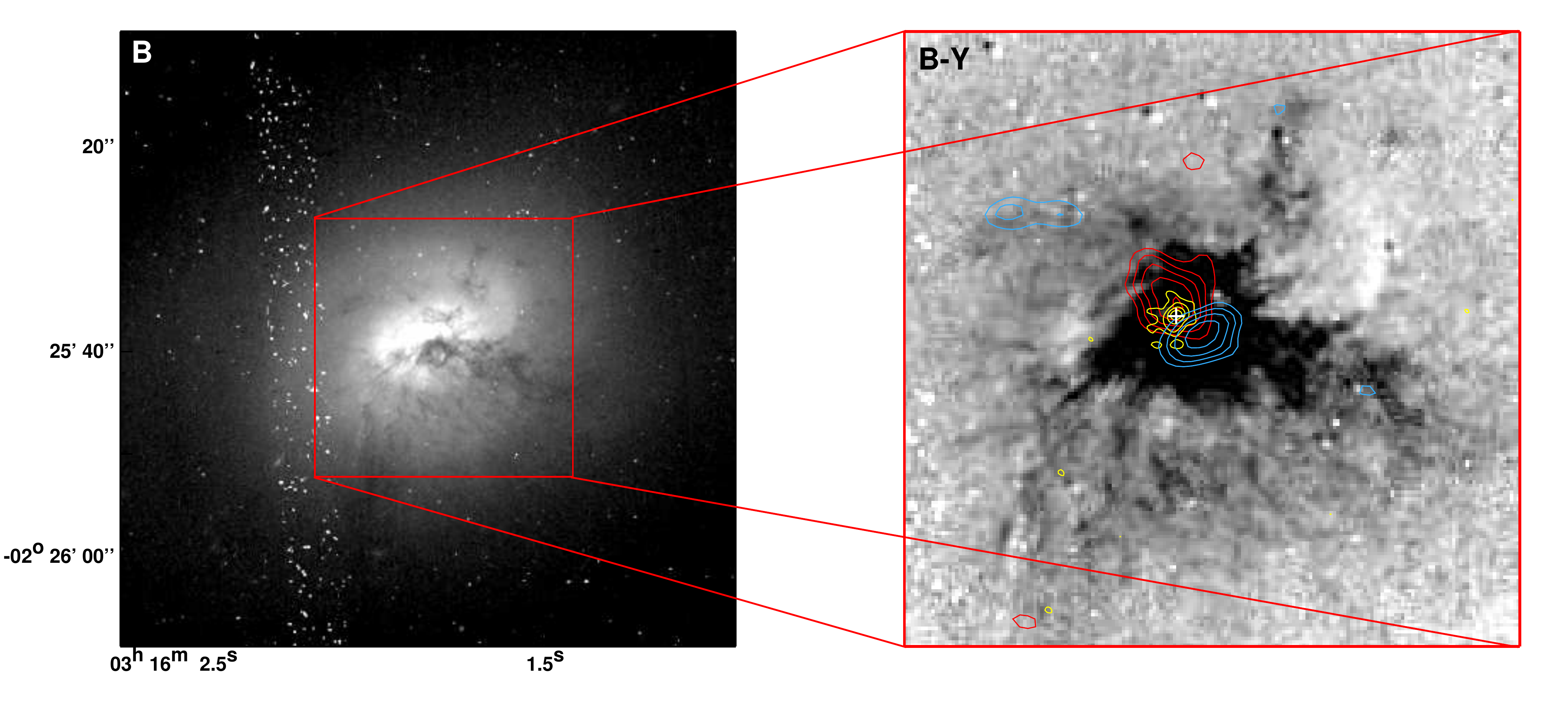}
\caption{{\bf (Left):} HST $B$-band image showing the extent of the dust extinction within NGC~1266.  {\bf (Right):}  Zoomed-in HST $B-Y$ color image highlighting the central dust features, with CO contours overlaid in red, blue and yellow.  The red and blue contours (SMA) correspond to the redshifted and blueshifted emission, respectively, while yellow contours (CARMA) correspond to the CO emission around $v_{\rm sys}$ only} \citep{alatalo+11}.  The white cross denotes the central position of the VLBA continuum emission.  A detailed analysis of the HST data will be presented in a future paper. 
\label{fig:co+dust}
\end{figure*}

NGC~1266 is a nearby (D~=~29.9~Mpc; \citealt{temi+09}) lenticular galaxy originally studied as part of the multiwavelength survey of 260 early-type galaxies currently being compiled by the \atlas\ collaboration \citep{cappellari+11}.  The presence of broad wings requiring a two-component Gaussian fit in the CO(2--1) spectrum from the Institut de Radioastronomie Millim\'{e}trique (IRAM) 30-m telescope was the first sign of the exceptional nature of NGC~1266 \citep{young+11}.  Follow-up millimeter observations performed with the Combined Array for Research in Millimeter Astronomy (CARMA) and the Submillimeter Array (SMA) suggested that NGC~1266 harbors a massive molecular outflow \citep{alatalo+11}.  SMA imaging revealed the presence of redshifted and blueshifted lobes, which are coincident with H$\alpha$ from the {\it Spitzer} Infrared Nearby Galaxy Survey (SINGS; \citealt{kennicutt+03}), 1.4~GHz continuum \citep{baan+06}, X-ray thermal bremsstrahlung \citep{alatalo+11}, and outflowing ionized and atomic gas emission \citep{davis+12}.  CO data suggest that $\sim10^9$~\msun\ of molecular gas is contained within the central 0.7$^{\prime \prime}$ (100~pc) of NGC~1266.  At least $\approx2.4\times10^7~{\rm M_\odot}$ of molecular gas\footnote{Assuming optically thin emission; see Alatalo et al.\ (2011) for further details.} extending out to a radius of at least 460~pc is involved in the outflow, although recent data of other molecular gas tracers indicate this mass may be much higher (\citealt{crocker+12, bayet+13}; Alatalo et al., in preparation).  The maximum velocities of the molecular, atomic, and ionized gas phases of the outflow are 480 km~s$^{-1}$, 500 km~s$^{-1}$ and 900 km~s$^{-1}$, respectively \citep{alatalo+11, davis+12}.  Davis et al.\ (2012) suggest that the discrepancy between the maximum velocities of the neutral and ionized gas components could be due to the destruction of the cold gas as it flows out of the galaxy or to observational sensitivity limitations.  

The velocity of the full width at half-maximum of the broad wing component of the molecular line emission is 177~km~s$^{-1}$ and corresponds to a dynamical time of 2.6~Myr, implying a mass outflow rate of at least $13~{\rm M_\odot}$~yr$^{-1}$.  At least 2.5\% of the wing emission exceeds the local escape velocity of $v_{\mathrm{esc}}$ = 340 km~s$^{-1}$ \citep{alatalo+11} and is thus enriching the intergalactic medium.  The molecular gas supply is thus expected to be exhausted in less than 100~Myr, effectively suppressing any significant future SF \citep{alatalo+11}.  This is supported by the recent analysis of Alatalo et al.\ (2013), in which a fit of stellar population models to the optical spectrum revealed that NGC 1266 harbors a poststarburst stellar population, with a single stellar population equivalent age of $\approx$1-2 Gyrs.

While compact nuclear starbursts can potentially drive outflows (e.g., Arp~220; \citealt{sakamoto+09}), Alatalo et al.\ (2011) showed that the current level of SF is insufficient to power the outflow in NGC~1266, suggesting that an additional energy source, such as a buried AGN, is responsible.  Sensitive, high-resolution, interferometric radio continuum observations have the potential to probe even heavily obscured environments and pinpoint emission from accreting massive black holes.  

Here we report the results of new Very Long Baseline Array (VLBA) 1.65~GHz continuum observations of the nucleus of NGC~1266.  We also report recent Hubble Space Telescope (HST) imaging and high spatial resolution Karl G. Jansky Very Large Array (VLA) H{\tt I} absorption observations.  In \S\ref{obs}, we describe the HST, VLA, and VLBA observations.  In \S\ref{results}, we discuss the results of our VLA and VLBA observations and explore possible origins for the emission.  In \S\ref{disc}, we discuss the implications of a low-level AGN origin for the VLBA-detected emission, re-evaluate the energetics of the radio continuum emission, and suggest that NGC~1266 should be regarded as a mildly Compton-thick system.  In \S\ref{conclu}, we summarize our results and suggest future directions for continued NGC~1266 studies.

\section{Observations and Data Reduction}
\label{obs}

\begin{table}
\caption{Summary of ACS HST Observations}
\label{tab:hst}
\begin{tabular*}{8.8cm}{c c c c c c}
\hline \hline
Obs ID & Instrument & Channel & Filter & Band & Exp. Time \\ 
& & & & & (seconds) \\
\hline
jbr704f6q & ACS & WFC & F814W & I  & 135 \\
jbr704f7q & ACS & WFC & F435W & B & 560 \\
jbr704f9q & ACS & WFC & F555W & V & 339 \\
jbr704fbq & ACS & WFC & F435W & B & 560 \\
jbr704fdq & ACS & WFC & F555W & V & 339 \\
\hline \hline
\end{tabular*}
\end{table}

\subsection{The Hubble Space Telescope}
Visible and infrared images of NGC~1266 were obtained with the Advanced Camera for Surveys (ACS) Wide Field Camera and the Wide Field Camera 3 (WFC3) instruments on the HST in December 2011 as part of Program~12525 over three orbits.   Table \ref{tab:hst} lists the identification, instrument, channel, filter, and exposure time for each dataset from the HST observations obtained with the ACS (see \citealt{alatalo+13} for a summary of the WFC3 data).  All ACS images are full frame and were processed with the standard reduction pipeline {\tt CALACS}.  In addition, destriping was applied to remove the characteristic horizontal noise pattern from the bias frames produced by one of the reference voltages supplied to the Wide Field Channel CCDs.  Cleaned images were coadded, registered, and scaled to a common pixel scale of 0.13 arcsec/pixel with the {\tt MULTIDRIZZLE} software package \citep{fruchter+09}.  Flux calibration was applied to the resulting drizzled\footnote{The ``drizzle" algorithm is used to perform a linear reconstruction of an image from undersampled, ``dithered" data \citep{fruchter+02}.  In this context, ``dithering"  refers to the series of spatially-offset exposures that comprise an observation with the goal of improving spatial/spectral resolution and increasing the signal-to-noise ratio \citep{koekemoer+02}.} images.  An HST $B$-band (F435W) image is presented in Figure~\ref{fig:co+dust}\footnote{The $Y$-band (F110W) data used to generate the $B-Y$ image in the right panel of Figure~\ref{fig:co+dust} is from the WFC3 instrument; see Alatalo et al.\ (2013) for a description of these data.}.  These images reveal filamentary structures in the central dust emission of NGC~1266, similar to those seen in other systems with outflows (e.g., NGC~3801; \citealt{das+05, hota+12}).  A detailed analysis of the full set of HST data from this program will be presented in a future paper (Deustua et al., in preparation).

\subsection{The Very Large Array}
NGC~1266 was observed with the NRAO\footnote{The National Radio Astronomy Observatory is a facility of the National Science Foundation operated under cooperative agreement by Associated Universities, Inc.} VLA in the A-configuration on August 14 and 15, 2012 (project ID: 11A-169) over 6 hours (4 hours on source).  The Wideband Interferometric Digital Architecture (WIDAR) correlator was configured with a bandwidth of 8~MHz divided into 256 channels centered on H{\tt I} at the systemic velocity of the galaxy (2170 km s$^{-1}$; \citealt{cappellari+11}).  These observations were phase-referenced to the calibrator J0339-0146 every 11 minutes with a switching angle of 6$\degr$.  The positional accuracy of our phase calibrator was $<$0.002$\arcsec$.  The calibrator 3C138 was used to set the amplitude scale to an accuracy of 3\% and calibrate the bandpass.  

\begin{figure*}[t!]
\centering
\includegraphics[trim=0.50in 0in 0in 0in, clip=true, height=3.0in]{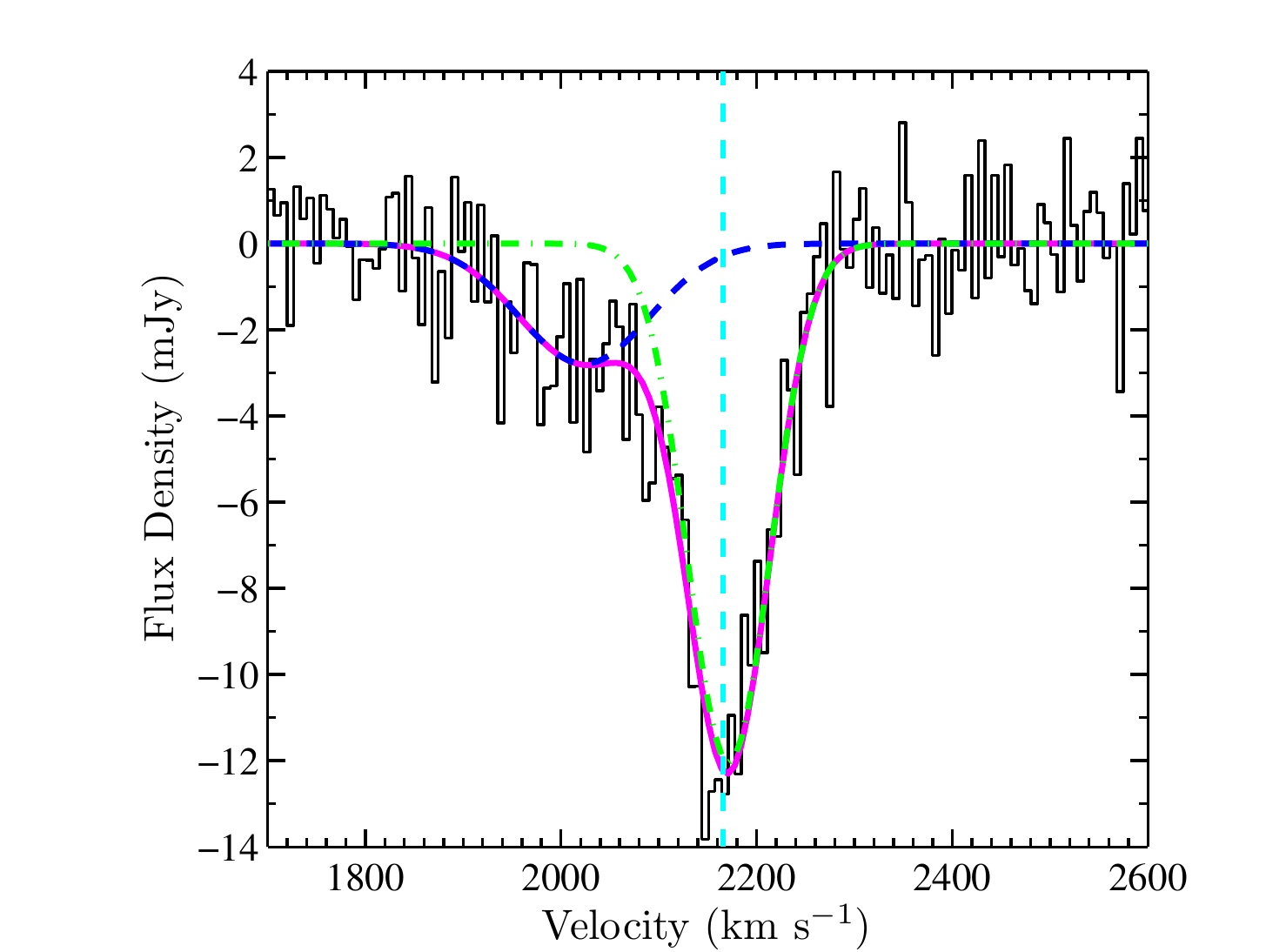} 
\caption{H{\tt I} absorption profile (black) in the central pixel of the data cube from the VLA A-configuration observations.  A double-Gaussian fit to the absorption profile is overlaid in magenta.  The two components consist of a shallow blueshifted outflowing component (blue dashed line) and a deeper absorption component near the systemic velocity (green dash-dot line).  The vertical dotted cyan line marks the systemic velocity of the Gaussian fit to the H{\tt I} absorption data of 2165 km s$^{-1}$.}
\label{fig:spectrum}
\end{figure*}

\begin{figure*}[t!]
\centering
\includegraphics[trim=0in 0in 0in 0in, clip=true, height=3.5in]{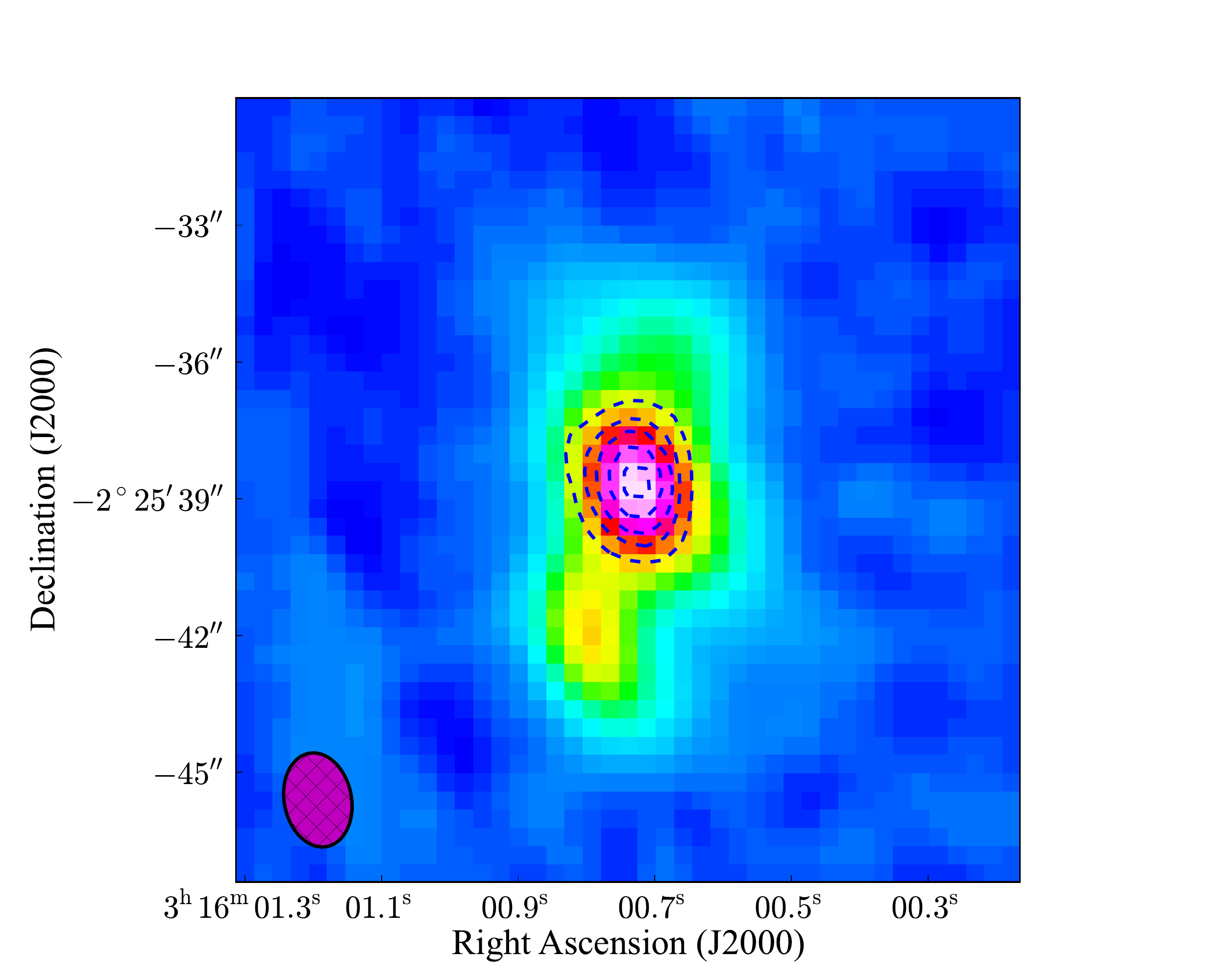} 
\caption{1.4~GHz continuum Stokes {\it I} image (background colorscale) with the continuum-subtracted integrated intensity (0th moment) H{\tt I} absorption contours overlaid in blue.  Negative contours are dashed.  The contour levels are [-850, -650, -450, -300, -150] mJy beam$^{-1}$ km s$^{-1}$.  From {\tt AIPS} JMFIT, the integrated flux density of the nuclear component of the underlying continuum emission coincident with the H{\tt I} absorption is 77.12 $\pm$ 2.47 mJy, with a peak flux density of 36.69 $\pm$ 1.14 mJy beam$^{-1}$.  Note that error bars were calculated by adding the standard VLA calibration error of 3\% in quadrature with the errors reported by JMFIT in {\tt AIPS}.  The spatial extent of the H{\tt I} absorption is 3.6$^{\prime \prime}$ (517.2~pc).  The total integrated flux density of the continuum emission including the extended lobes is 108.30 $\pm$ 3.25 mJy and spans 12$^{\prime \prime}$ (1.7~kpc).  The beam is shown in purple in the lower left corner of the figure and has major $\times$ minor axis dimensions of 2.08$^{\prime \prime}$ $\times$ 1.48$^{\prime \prime}$.}
\label{fig:vla_map}
\end{figure*}

We flagged, calibrated, and imaged the data with the December 31, 2011 release of the Astronomical Image Processing System ({\tt AIPS}) using standard procedures.  The rms noise of the clean\footnote{In radio astronomy, a ``clean" image is one in which the ``dirty" beam, or point spread function, has been deconvolved from the true image of the source.}, continuum-subtracted cube was 0.7 mJy beam$^{-1}$ per channel.  The synthesized beam dimensions are 2.08$^{\prime \prime}$ $\times$ 1.48$^{\prime \prime}$ and the channel width is 31.25~kHz = 6.6 km s$^{-1}$.  No H{\tt I} emission was detected, however continuum emission and H{\tt I} absorption were detected.  We extracted the absorption spectrum from the clean cube using the {\tt AIPS} task ISPEC (Figure~\ref{fig:spectrum} and Table~\ref{tab:vla_HI}) and produced an integrated intensity (0th moment) map of the H{\tt I} absorption using the MOMNT task (Figure~\ref{fig:vla_map}).  The VLA continuum and H{\tt I} absorption data are further discussed in Section~\ref{results:vla}.

We also re-examined archival continuum VLA data at 1.4 and 5~GHz.  These archival observations were carried out on December 14, 1992 in the A-configuration as part of observing program AB0660.  We reduced the data from these observations in {\tt AIPS} following standard procedures.  Although these data have been presented in the literature \citep{baan+06, alatalo+11, davis+12}, we include a brief re-analysis (Section~\ref{results:vla_archival}) here for the sake of completeness and clarity and provide contour maps (Figures~\ref{fig:vla_archival} and \ref{fig:vla_archival2}), flux density measurements (Table~\ref{tab:vla_archival}), and matched spatial resolution radio spectral index measurements (Section~\ref{results:vla_archival}).

\subsection{The Very Long Baseline Array}
NGC~1266 and calibrators were observed with all 10 antennas of the VLBA on October 5, 2011 over 8 hours (project ID: BA99).  The observations were centered at a frequency of 1.6564~GHz with a bandwidth of 16~MHz.  All VLBA data were correlated using the new DiFX software correlator \citep{deller+11} in dual polarization mode.  Due to the expected faintness of the target source, we utilized phase referencing with a switching angle of $2^\circ$.  Two-minute scans on the phase, rate, and delay calibrator, J0318-0029, preceded and followed three-minute scans of NGC~1266.  The calibrators J0238+1636 and J0312+0133 were also observed as coherence-check sources.  Data editing and calibration were performed with {\tt AIPS} using the standard VLBA procedures.

The {\tt AIPS} task IMAGR was used to form and deconvolve the image of the Stokes {\em I} emission of NGC~1266 (Figure~\ref{fig:vlba}b).  Visibility data were uniformly weighted in order to minimize artifacts due to residual radio frequency interference.  We used the {\tt AIPS} task JMFIT to fit a single elliptical Gaussian component to the emission to determine its flux density, extent, and position.  Observational parameters are summarized in Table~\ref{vlba} and a VLBA continuum image with contours is shown in Figure \ref{fig:vlba}b.  The milliarcsecond-scale ($\sim1$~pc) resolution observations of NGC~1266 reveal unresolved continuum emission located within the central molecular disk (Figure~\ref{fig:co+dust}b) \citep{alatalo+11}, at the kinematic center of the galaxy \citep{davis+12}.  These data are further discussed in Section~\ref{results:vlba}.

\begin{table*}
\caption{Summary of VLA H{\tt I} Absorption Gaussian Fit Parameters}
\centering
\label{tab:vla_HI}
\begin{tabular*}{12.5cm}{ccccccc}
\hline \hline
 Component & V$_{\mathrm{rel}}$ &  $\Delta \mathrm{v}_{\mathrm{FWHM}}$ & $S_{\mathrm{abs}}$ & $\tau_{\mathrm{peak}}$ & $N_{\mathrm{H{\tt I}}}$  & $M_{\mathrm{H{\tt I}}}$\\ 
                       & (km s$^{-1}$)      & (km s$^{-1}$)  &  (mJy beam$^{-1}$)  &    &  (10$^{21}$ cm$^{-2}$) & (10$^6$ M$_{\odot}$)\\ 
(1) & (2) & (3) & (4) & (5) & (6) & (7)\\ 
\hline
Blueshifted         & -141.7  & 157.5 & -2.79 & 0.08 & 2.41 & 1.41 \\
Deep Systemic  & 5.8         & 99.1  & -12.05 & 0.40 & 7.62  & 4.47 \\
\hline \hline
\end{tabular*}
\tablecomments{Col.\ (1): Gaussian component name.  Col.\ (2): Central velocity relative to the systemic velocity of 2165 km s$^{-1}$.  Col.\ (3): Gaussian width (full velocity at half maximum).  Col.\ (4): Peak absorption flux density (or, equivalently, the peak of the corresponding Gaussian distribution).  Col.\ (5): Peak optical depth, $\tau_{\mathrm{peak}} \equiv \ln  \left(  S_{\mathrm{cont}}/(S_{\mathrm{cont}} + S_{\mathrm{abs}}) \right)$, where $S_{\mathrm{cont}}$ is the peak continuum flux density in mJy beam$^{-1}$ (see Figure~\ref{fig:vla_map}).  Col.\ (6): Column density, $N_{\mathrm{H{\tt I}}} \equiv 1.823 \times 10^{18}  \, T_{\mathrm{s}} \int \tau (v) \, dv$ cm$^{-2}$, where $T_{\mathrm{s}}$ is the spin temperature, $\tau (v)$ is the optical depth, and $dv$ is the velocity in km s$^{-1}$.  After fitting a Gaussian to the absorption spectrum, $\int \tau (v) dv = 1.06 \, \tau_{\mathrm{peak}} \, \Delta \mathrm{v}_{\mathrm{FWHM}}$.  Col.\ (7): Mass of H{\tt I}, $M_{\mathrm{H{\tt I}}} \equiv M_{\mathrm{H}} \, N_{\mathrm{H{\tt I}}} \, A$, where $M_{\mathrm{H}}$ is the mass of a Hydrogen atom and $A$ is the presumed area of the absorbing H{\tt I} cloud.}
\end{table*}

\section{Results}
\label{results}
\subsection{New High-resolution 21cm Absorption Data}
\label{results:vla}
A VLA radio continuum image overlaid with H{\tt I} absorption contours is shown in Figure~\ref{fig:vla_map}.  These VLA A-configuration H{\tt I} absorption data were obtained as follow-up to a previous H{\tt I} absorption project carried out in the VLA D-configuration in 2010 and presented in Alatalo et al.\ (2011).  Like the lower-resolution, D-configuration data, the absorption detected in the A-configuration observations is unresolved (diameter $\approx$ 302~pc).  However, the improved spatial resolution of the VLA A-configuration data ($\theta_{\mathrm{FWHM}}$ = 2.08$^{\prime \prime} \times 1.48^{\prime \prime}$) shows that the absorption is co-located with the compact, nuclear, 1.4~GHz continuum emission detected by the VLA at matched spatial resolution.  Since the morphology of the 1.4~GHz VLA continuum emission is highly suggestive of an AGN (\citealt{alatalo+11, davis+12}; Section~\ref{results:vla_archival} of this work), we consider the most likely origin of the underlying continuum emission absorbed by the H{\tt I} to be an AGN.

Despite the likelihood that radio continuum emission from the putative AGN is causing the H{\tt I} absorption, the radio ``lobes" evident in the 1.4~GHz VLA continuum image do not appear to be associated with any H{\tt I} absorption.  This could be due to the fact that the {\tt HI} phase of the outflow is truly localized very close to the central black hole.  Another possibility is that the underlying continuum emission in the lobes is too weak for any extended {\tt HI} gas to be visible in absorption.  The radio lobes have peak flux densities of 10.06 and 8.81 mJy beam$^{-1}$ for the southern and northern lobe, respectively.  Assuming a 3$\sigma$ upper limit of 2.1 mJy beam$^{-1}$ per channel in the continuum-subtracted H{\tt I} absorption cube, the upper limit to the peak optical depth in each lobe is 0.23 (southern lobe) and 0.27 (northern lobe).  Thus, our new VLA data may simply not be sensitive enough to detect a diffuse, extended component of the H{\tt I} absorption.

The presence of a faint, extended H{\tt I} absorption component that has not yet been detected is supported by the spatial extent of NaD absorption, which can be used as a proxy for H{\tt I} column density \citep{cardelli+89, lehnert+11, davis+12}.  The NaD absorption is more extended than the H{\tt I} absorption (see Figure~18c, \citealt{davis+12}), and has a projected diameter of 3.25~$\pm$ 0.25$^{\prime \prime}$ (470 $\pm$ 35~pc).  This again suggests that although extended H{\tt I} absorption may be present, the radio continuum is likely too weak to allow us to detect it at larger radii.

As shown in Figure~\ref{fig:spectrum}, the H{\tt I} absorption spectrum can be modeled with a double Gaussian composed of a deep component near the systemic velocity and a shallower blueshifted component associated with the atomic gas portion of the outflow.  The Gaussian fit properties and derived parameters are summarized in Table~\ref{tab:vla_HI}.  Since the absorbing gas must be in front of the continuum source, the fact that we only see a blueshifted component in the H{\tt I} absorption spectrum indicates that the gas is indeed outflowing \citep{morganti+05b, alatalo+11}.

Assuming a spin temperature $T_{\mathrm{s}}$ = 100~K, the total column density of the H{\tt I} is $N_{\mathrm{H{\tt I}}}$ = 1.00$\times 10^{22}$ cm$^{-2}$.  The deep systemic and blueshifted components contribute column densities of  $N_{\mathrm{H{\tt I}}}$ = 7.62$\times 10^{21}$ and 2.41$\times 10^{21}$ cm$^{-2}$, respectively.  Since the spin temperature may be significantly higher in reality, all of our parameters derived from the H{\tt I} absorption data should be regarded as lower limits.  This is especially true if the gas is physically located near the AGN where it can be easily heated.  These H{\tt I} column density measurements are consistent with those in Alatalo et al.\ (2011) from the lower spatial resolution H{\tt I} absorption data, as well as the NaD absorption-derived H{\tt I} column density estimates calculated using the two methods described in detail in Davis et al.\ (2012).

The mass of the absorbing H{\tt I} column can be calculated by integrating the derived column density over the presumed spatial extent of the H{\tt I}\footnote{Note that the ``standard" equation for the mass of H{\tt I} detected in emission ($M_{\mathrm{H{\tt I}}} = 2.36 \times 10^5  D^2  \int S(v) dv$ M$_{\odot}$, where $D$ is the distance in Mpc and $\int S(v) dv$ is the line flux in units of Jy km s$^{-1}$) does not apply in the case of H{\tt I} absorption since the strength of the H{\tt I} absorption depends on the strength of the underlying continuum emission.}.  Assuming that the absorbing H{\tt I} cloud occupies the full extent of our beam ($\theta_{\mathrm{FWHM}} = 2.08^{\prime \prime}$ = 302~pc), we find a total H{\tt I} mass of $M_{\mathrm{H{\tt I}}}$ = 5.88 $\times$ 10$^6$ M$_{\odot}$.  This total H{\tt I} mass can be broken down into contributions from the deep systemic and blueshifted components of $M_{\mathrm{H{\tt I}}}$ = 4.47 $\times$ 10$^6$ and 1.41 $\times$ 10$^6$ M$_{\odot}$, respectively.  These values are consistent with the H{\tt I} absorption mass estimates provided in Alatalo et al.\ (2011).

As Alatalo et al.\ (2011) pointed out, the H{\tt I} absorption properties of NGC~1266 are similar to those found in radio galaxies potentially harboring outflows, albeit on a smaller scale.  These more powerful AGNs, (e.g., FRI/FRII radio galaxies; \citealt{fanaroff+74}) have jets typically spanning kpc-scales (though some are known to have extents larger than 1~Mpc; \citealt{koziel+11}).  The neutral outflows of these radio galaxies often exhibit spatially extended H{\tt I} absorption with spectral full widths at zero intensity (FWZI) ranging from 600 to nearly 2000 km s$^{-1}$, and relative velocities in excess of 1000 km s$^{-1}$ (e.g., \citealt{morganti+03, morganti+05b}).  The high H{\tt I} absorption velocities, location of the H{\tt I} absorption, and evidence for strong shocks in ionized gas studies in these objects have been cited as evidence of jet-accelerated neutral outflows (e.g., \citealt{morganti+05a, morganti+07}).

The FWHM of the blueshifted component of the H{\tt I} absorption in NGC~1266 is 157.5 km s$^{-1}$ (FWZI $\approx$360 km s$^{-1}$) and the relative velocity is $-141.7$ km s$^{-1}$.  The relative velocity derived from the H{\tt I} absorption is less than the typical NaD absorption blueshifted velocity of $-250$ km s$^{-1}$ and maximum velocity of $-500$ km s$^{-1}$ \citep{davis+12}.  It is possible that the H{\tt I} outflow extends to higher velocities more similar to those indicated by the NaD absorption data, but our VLA data do not have the necessary sensitivity to detect this higher-velocity component.  Another possibility is that the relatively low outflow velocity and lack of extended H{\tt I} absorption in NGC~1266 is an indication that the putative AGN jet in this galaxy is too weak to drive the outflow.  

Recent studies of the H{\tt I} absorption properties of nearby, lower-power, young, compact radio sources, such as ``gigahertz peaked-spectrum" (GPS) and ``compact steep-spectrum" (CSS) objects \citep{odea+98}, provide another point of comparison for the H{\tt I} absorption properties of NGC~1266.  GPS and CSS objects have dynamical ages of $\sim$10$^{2}-$10$^4$ yr and $>$10$^5$ yr, respectively, and are believed to be the youthful versions of more powerful, classical radio galaxies with dynamical ages over 10$^8$ yr \citep{odea+98, murgia+99}.  With typical jet extents of less than 1~kpc for GPS sources and 1 to 20~kpc for CSS sources, they are also considerably more compact than radio AGNs in the FRI/FRII regime, and more similar to the $\approx$1.7~kpc extent of the VLA 1.4~GHz continuum emission in NGC~1266\footnote{We are not necessarily suggesting that NGC~1266 should be classified as a young radio source; we discuss young radio sources here for comparative purposes only.} (Figure~\ref{fig:vla_map}).  

Despite their relatively compact jet morphologies compared to classical radio galaxies, H{\tt I} absorption studies have suggested that the radio jets in these young GPS and CSS sources may be capable of producing jet-cloud interactions that can significantly impact their host galaxies (e.g., \citealt{snellen+04, gupta+06, chandola+11}).  The sample of young radio sources compiled by Chandola et al.\ (2011) provides an interesting comparison to NGC~1266.  Although the continuum properties of the young radio sources are all at least an order of magnitude larger than those found in NGC~1266, they are still more similar to NGC~1266 than powerful radio galaxies with massive jets.  The sources detected in H{\tt I} absorption in Chandola et al.\ (2011) have blueshifted component velocities ranging from $-0.9$ to $-256.1$ km s$^{-1}$.  Despite these relatively low outflow velocities, the authors cite a variety of multiwavelength evidence in support of the possibility that jet-cloud interactions can produce at least some of the H{\tt I} absorption features observed in these sources.  This topic remains controversial, however.  Continued studies of samples of lower power, compact radio sources with blueshifted H{\tt I} absorption components will be necessary to help place constraints on jet-driven AGN feedback in local galaxies.
 
\subsection{Archival VLA Continuum Data}
\label{results:vla_archival}
\begin{figure*}[t!]
\centering
\includegraphics[width=3in]{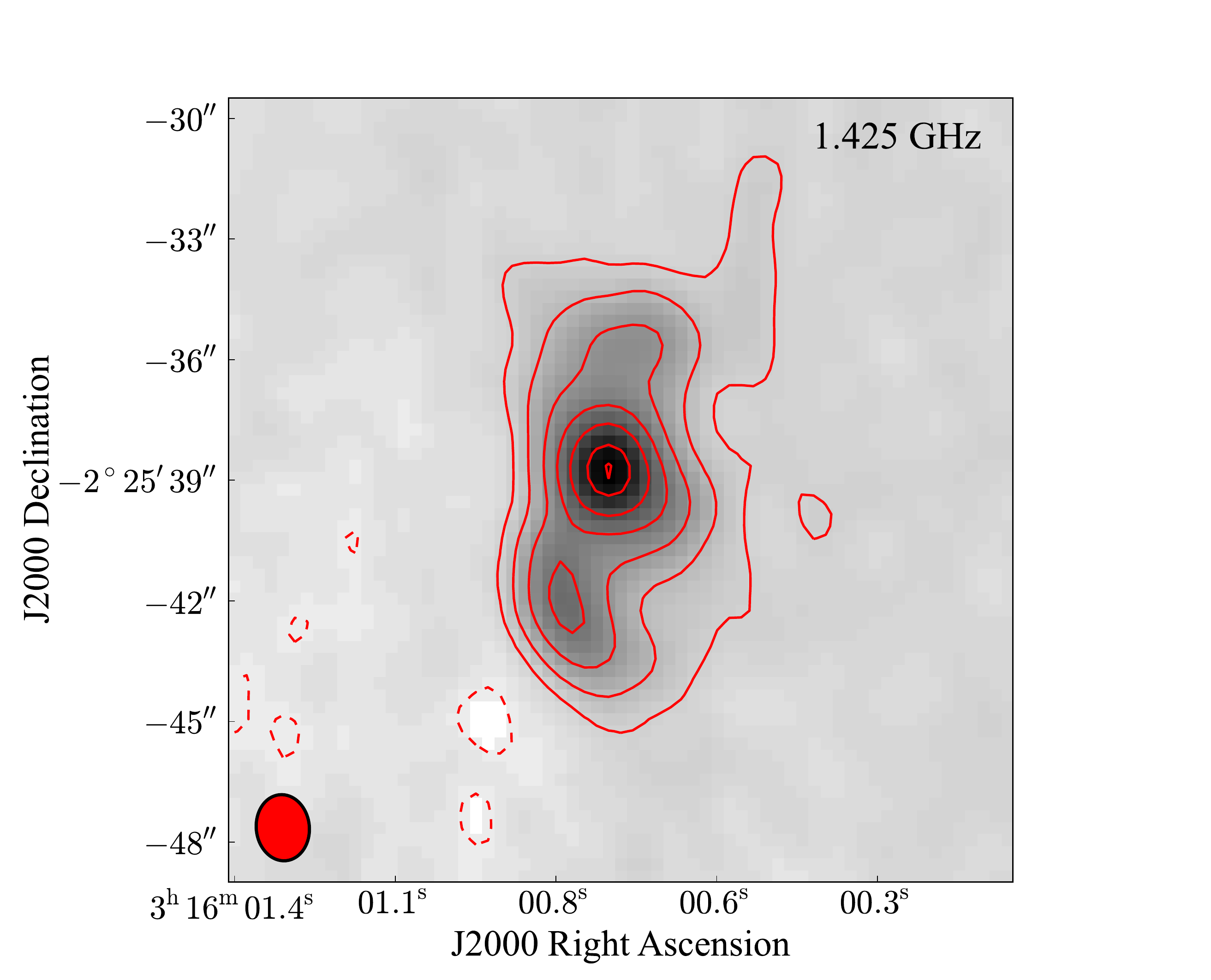}

\includegraphics[width=3in]{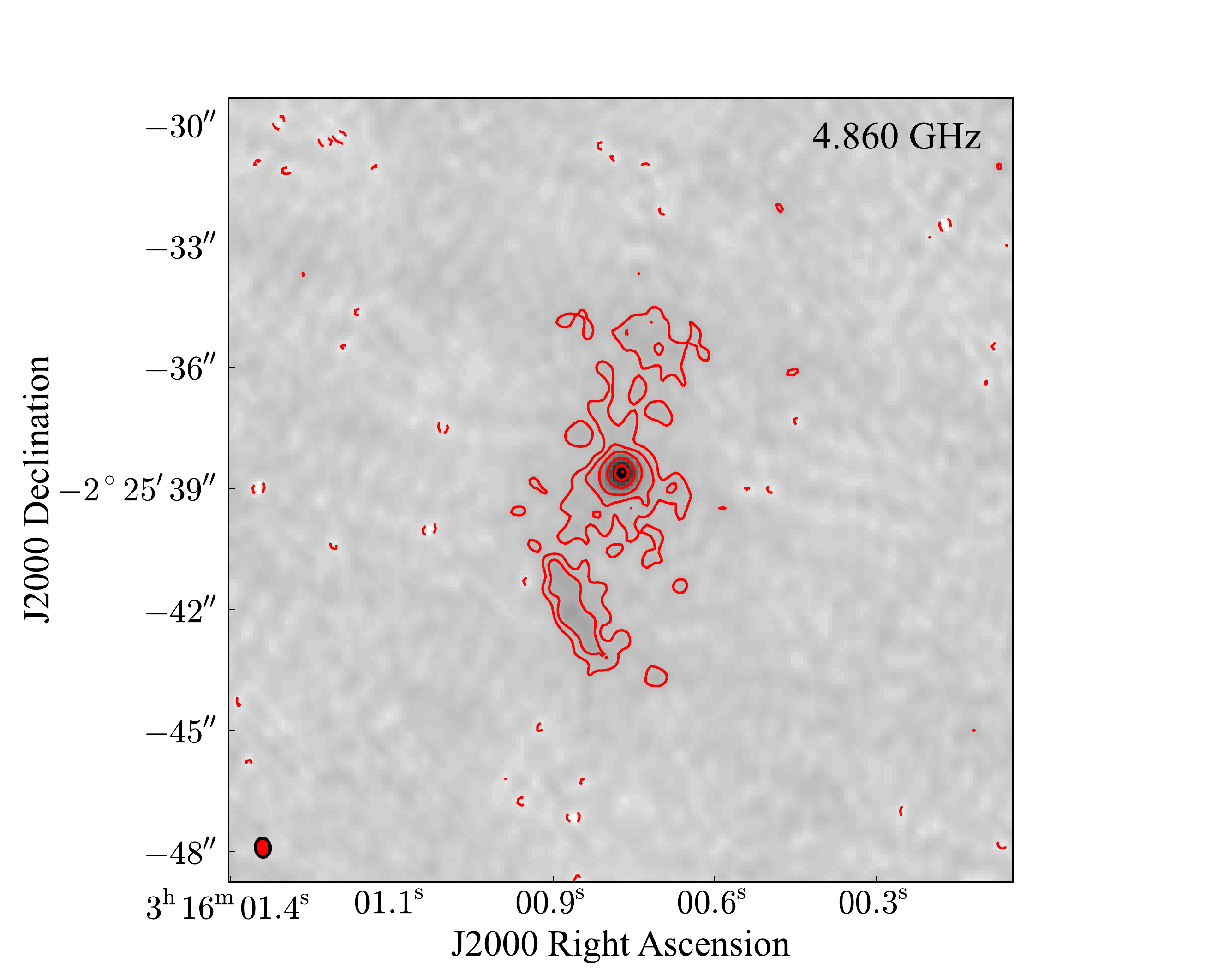}
\includegraphics[width=3in]{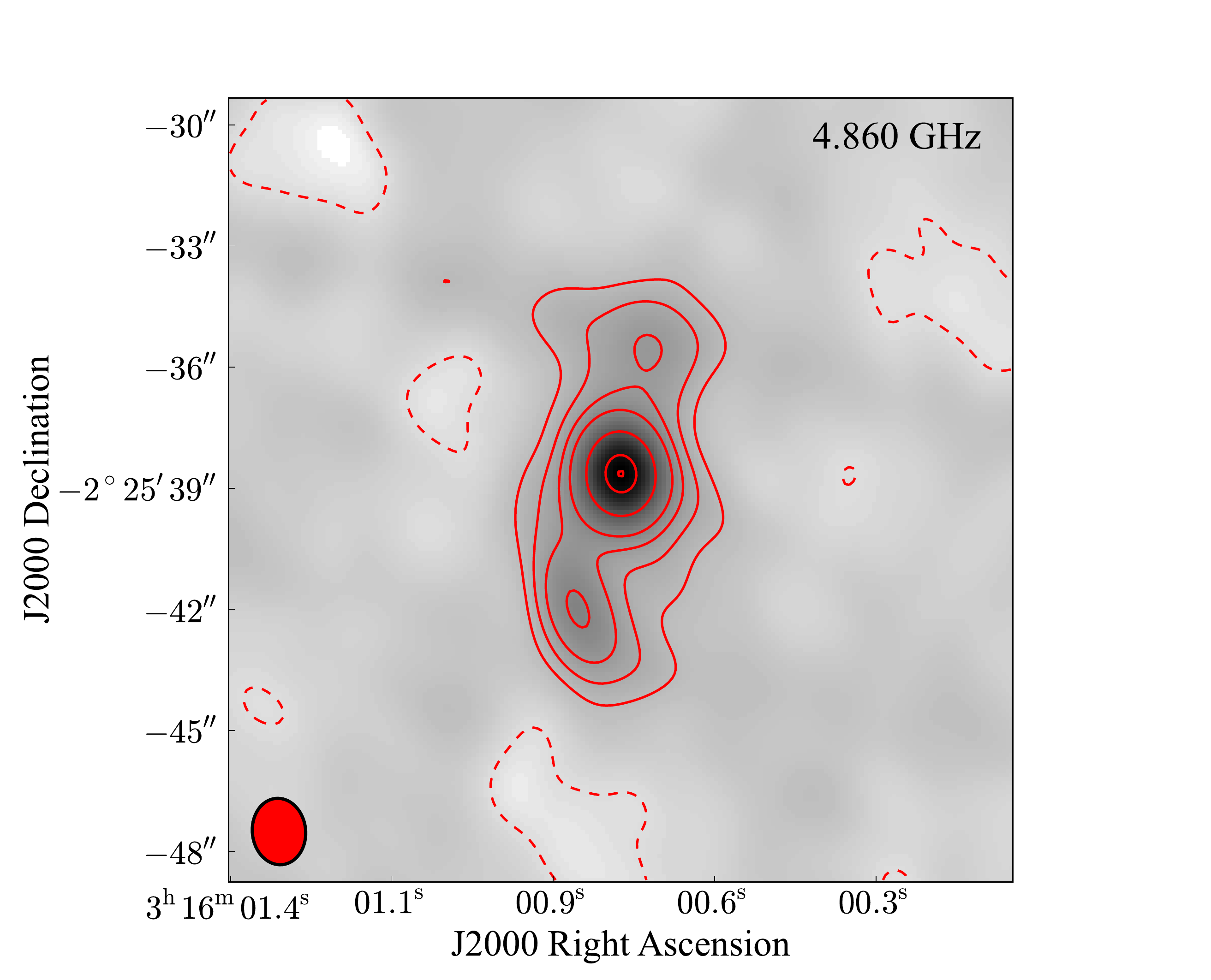}
\caption{NGC~1266 archival VLA radio continuum images with contours at 1.4 and 5~GHz.  For the 1.4~GHz image (top), the relative contour levels are [-3, 3, 12, 30, 58, 102, 180, 240] and the unit contour level is 0.15 mJy beam$^{-1}$.  Two 5~GHz images are shown (bottom row).  The image on the left is shown with the native spatial resolution and the image on the right has been smoothed to match the resolution of the 1.4~GHz image.  The relative contour levels for the native resolution 5~GHz image are [-3, 3, 6, 15, 50, 136, 196] and the unit contour is 0.07 mJy beam$^{-1}$.  For the smoothed 5~GHz image, the relative contour levels are [-3, 3, 6, 10, 16, 36, 72, 86] and the unit contour level is 0.25 mJy beam$^{-1}$.  The unit contour levels represent the rms noise in each image.    The VLA beam is the filled ellipse shown at the lower left in each image, and the clean beam parameters are given in Table~\ref{tab:vla_archival}.}
\label{fig:vla_archival}
\end{figure*}

\begin{figure*}[t!]
\centering
\includegraphics[trim=0in 0in 0in 0in, clip=true, height=3.0in]{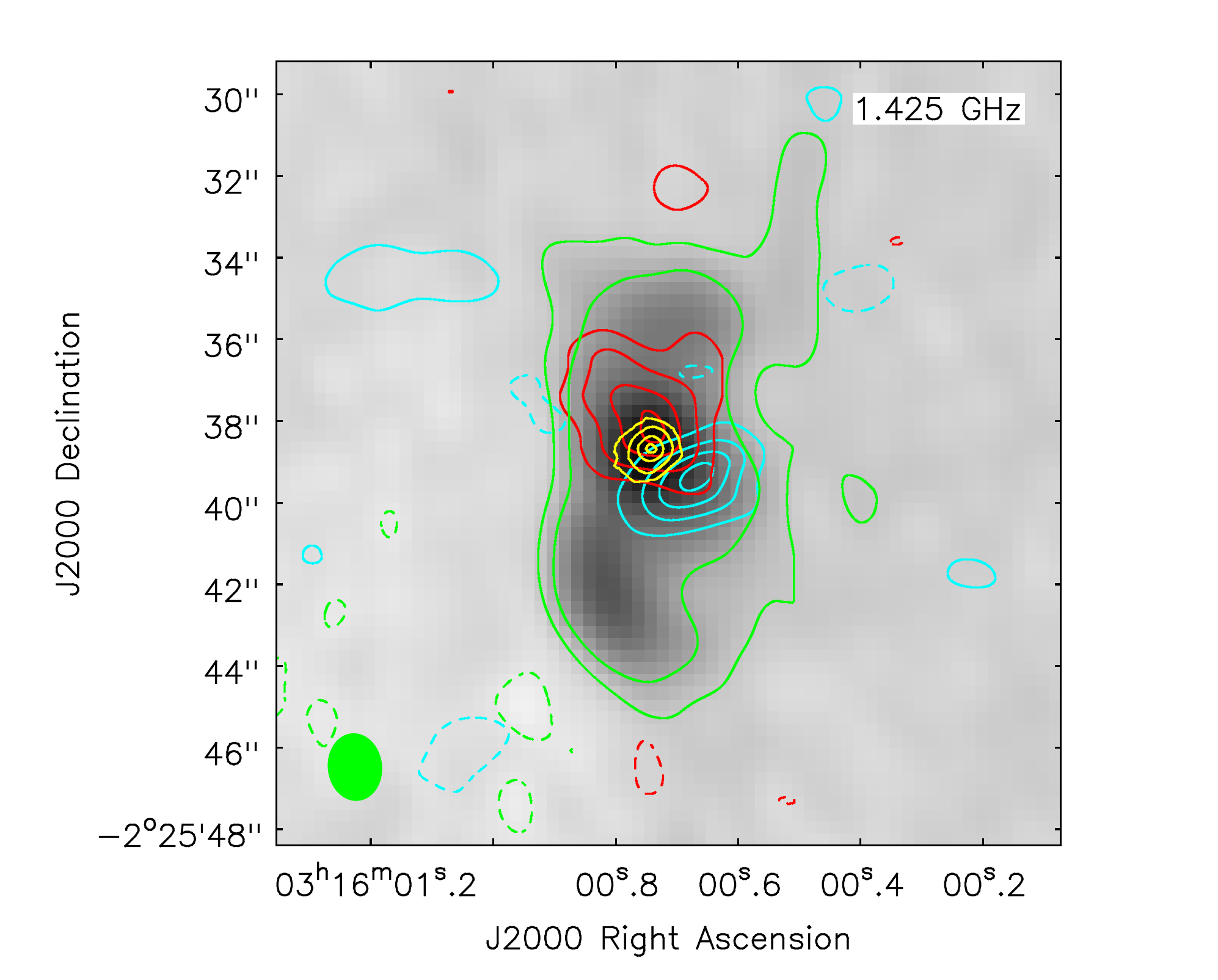}
\caption{NGC~1266 archival 1.4~GHz VLA radio continuum image (greyscale and green contours; central observing frequency shown in the upper right) with CO contours overlaid.  The VLA beam is the filled green ellipse shown in the lower left corner of the figure.  Dashed contours are negative.  The two green contours are the archival VLA 1.4~GHz 3$\sigma$ and 12$\sigma$ contours from Figure~\ref{fig:vla_archival}, shown here to help highlight the extent of the low surface brightness continuum emission of the lobes.  The yellow contours represent the integrated intensity map of the CARMA CO(1--0) emission near $v_{\mathrm{sys}}$.  The relative contour levels are [-3, 3, 6, 11, 15] and the unit contour level is 1.1 Jy beam$^{-1}$ km s$^{-1}$.  The red and cyan contours represent the SMA CO(2--1) integrated intensity maps of the redshifted and blueshifted spectral components, respectively.  The relative contours for the redshifted component are [-3, 3, 5, 7, 8], with a unit contour level of 0.85 Jy beam$^{-1}$ km s$^{-1}$.  For the blueshifted component, the relative contour levels are [-3, 3, 5, 6.25, 7.1], with a unit contour level of 0.87 Jy beam$^{-1}$ km s$^{-1}$. The spatial resolution of the CO data is 0.5$^{\prime \prime}$ $\times$ 0.4$^{\prime \prime}$ for the CARMA data and 2.3$^{\prime \prime}$ $\times$ 2.0$^{\prime \prime}$ for the SMA data.  For detailed information on the CARMA and SMA data, see Alatalo et al.\ (2011).}
\label{fig:vla_archival2}
\end{figure*}

The archival VLA data reveal a bright, compact core surrounded by lobes of diffuse continuum emission to the north and south (Figures~\ref{fig:vla_archival} and \ref{fig:vla_archival2}).  Alatalo et al.\ (2011) suggested that the extended portions of the radio emission could have a jet origin.  In a follow-up paper, Davis et al.\ (2012) reported that the extended radio emission correlates well with ionized and atomic gas features, in line with expectations of jet propagation in a dense environment.  Furthermore, the spatial coincidence between the edge of the southern portion of the extended radio emission and shock features (from integral field unit spectroscopic optical line ratio maps) suggests that the radio jet may be interacting with the ISM \citep{davis+12}.

Recent studies of other galaxies hosting molecular outflows have suggested that, for molecular gas entrainment by a radio jet, both the energy and geometry of the jet and the molecular gas should be similar (e.g., M51; \citealt{matsushita+04, matsushita+07}; NGC~3801; \citealt{das+05}).  In this section, we compare the geometry of the molecular outflow with that of the radio jet; jet energetics are discussed in Section~\ref{energy}.  

In Figure~\ref{fig:vla_archival2}, we have overlaid the CARMA CO(1--0) data around $v_{\mathrm{sys}}$ and the SMA redshifted and blueshifted CO(2--1) data on the 1.4~GHz archival VLA continuum data.  The central molecular CO(1--0) disk appears to be spatially coincident with the core of the radio continuum emission.  However, the CO(2--1) redshifted and blueshifted components are much more compact than the extended radio continuum  emission.  Similar to the possible explanations provided for the compact nature of the H{\tt I} absorption in Section~\ref{results:vla}, the compactness of the molecular gas relative to the VLA radio continuum emission could be due to the destruction of this cold gas as it flows out of the galaxy or sensitivity limitations.  It is also possible that radiation-driving powered by an AGN wind is responsible for the molecular outflow.  Upcoming detailed molecular gas studies with ALMA, which can isolate shock tracers expected to exist in photon-dominated regions, may help to definitively determine the viability of a radiative wind as the driver of the outflow.

Additionally, the position angle of the molecular gas appears offset from that of the extended radio continuum emission in Figure~\ref{fig:vla_archival2}, similar to the offset observed in Figure~10 of Davis et al.\ (2012) between the NaD absorption and the ionized gas.  Davis et al.\ (2012) suggested that this discrepancy in position angle could be due to projection effects, deflection of the jet due to the dense ISM, jet precession, or different driving mechanisms for different components of the outflow.  The southern lobe of the 1.4~GHz continuum emission does show some curvature in the direction of the blueshifted molecular gas (Figure~\ref{fig:vla_archival2}, as highlighted by the 12$\sigma$ contour of the VLA continuum emission), suggesting that deflection of the jet by the ISM is a plausible scenario for the position angle offset between the radio jet and the molecular gas.  A more detailed analysis of the properties of the redshifted and blueshifted molecular outflow components would allow us to evaluate whether or not NGC~1266 may host jet-entrained molecular gas.  Such an analysis awaits a future study with ALMA, which can provide significantly improved spatial and spectral resolution as well as higher sensitivity.

\begin{table*}[t!]
\caption{Summary of Archival VLA Continuum Observations}
\centering
\label{tab:vla_archival}
\begin{tabular*}{9.5cm}{cccccc}
\hline \hline
  & \multicolumn{2}{c}{Beam Parameters} & &\multicolumn{2}{c}{Source Parameters}  \\ 
\cline{2-3} \cline{5-6} \\
 (Frequency) & ($\theta_{M} \times \theta_{m}$) & (P.A.) & & ($S_{\mathrm{tot}}$) & ($S_{\mathrm{core}}$)\\ 
(GHz)  & ($^{\prime \prime}$) & (deg) & &  (mJy) & (mJy)\\ 
(1) & (2) & (3) & & (4) & (5)\\ 
\hline
1.4  & 1.65 $\times$ 1.32 & 5.9 &  &   90.34 $\pm$ 0.83 & 70.53 $\pm$ 0.60\\
5  & 0.47 $\times$ 0.37 & 6.7 & & 20.33 $\pm$ 0.19 & 20.30 $\pm$ 0.19 \\
5  & 1.65 $\times$ 1.32 & 5.9 &  &  34.05 $\pm$ 0.96 & 31.43 $\pm$ 0.88\\
\hline \hline
\end{tabular*}
\tablecomments{Col.\ (1): Central frequency.  Col.\ (2): Clean beam major $\times$ minor axis.  Col.\ (3): Clean beam position angle. Col.\ (4): Total integrated flux density of the core plus the extended emission with the associated error.  The error listed is the sum of the error reported by JMFIT in {\tt AIPS} and the standard 3\% VLA calibration error, added in quadrature.  Col.\ (5): Same as Col.\ (4) but for the integrated core flux density only.}
\end{table*}

Integrated flux densities for the total extent of the emission and for the core only at 1.4 and 5~GHz are provided in Table~\ref{tab:vla_archival}.  Since these archival data were obtained nearly simultaneously, they offer a reliable means of calculating the radio spectral index.  After smoothing the 5~GHz data to match the spatial resolution of the 1.4~GHz data, we find a radio spectral index (where $S \varpropto \mathrm{\nu}^{\mathrm{\alpha}}$) for the core emission of $\alpha_{\mathrm{core}} = -0.66 \pm 0.02$.  When the extended portions of the radio emission are included, the spectral index steepens a bit to $\alpha_{\mathrm{tot}} = -0.79 \pm 0.02$.  However, we caution that the extent of the 1.4~GHz emission from the VLA A-configuration ($\approx$12$^{\prime \prime}$) is larger than the largest angular scale detectable by the 5~GHz VLA data from the same configuration ($\approx$9.4$^{\prime \prime}$).  Therefore, the 5~GHz data may have resolved-out some portion of the extended, low surface brightness emission that is measured in the 1.4~GHz data, which could artificially steepen the spectral index\footnote{Our spectral index measurements of the core emission should not be affected by these differences in uv-coverage.} in the outer parts.  However, if the spectral steepening in the lobes is real, this would be consistent with spectral index studies of other systems with extended radio emission due to AGN activity (e.g., \citealt{ho+01, giroletti+05, laing+11}).

\begin{table}[t!]
\caption{VLBA Observational Parameters}
\label{vlba}
\begin{tabular*}{8.5cm}{c c l}
\hline \hline
Frequency & (MHz) & 1656$^a$\\
rms noise & ($\mu$Jy beam$^{-1}$) & 42$^b$\\
\hline
\hspace{7mm}{\bf Beam} & {\bf Parameters}\hspace{7mm} & \\
\hline
$\theta_M\times\theta_m$ & (mas) & $9.75\times4.31^c$\\
P.A. & (deg) & -0.39$^d$\\
\hline
\hspace{5mm}{\bf Source} & {\bf Parameters}\hspace{7mm} & \\
\hline
$\theta_M\times\theta_m$ & (mas) & $7.98\pm1.46\times6.19\pm0.96^e$\\
P.A. & (deg) & $10.39\pm40.88^f$\\
$M\times m$ & (pc) & $1.16\pm0.21\times0.89\pm0.14^g$ \\
$S$ & (mJy) & $1.38\pm0.14^h$ \\
log($P_{\rm rad}$) & (W Hz$^{-1}$) & 20.17$^i$\\
$T_{\rm b}$ & (K) & $1.5\times10^{7~j}$ \\
$\alpha_{\rm J2000}$ & (hms) & $03^{\rm h}16^{\rm m}00.742^{\rm s~k}$ \\
$\delta_{\rm J2000}$ & (dms) & $-02^\circ25'38.66''^{~l}$ \\
\hline \hline
\end{tabular*} \\
\vskip 1mm
\noindent$^a$ Central frequency. \\
\noindent$^b$ Average rms noise in image.\\
\noindent$^c$ Clean beam major $\times$ minor axis.\\
\noindent$^d$ Clean beam position angle.\\
\noindent$^e$ Angular dimensions (major $\times$ minor axis) and error bars from JMFIT in {\tt AIPS}.\\
\noindent$^f$ Position angle of deconvolved emission from JMFIT in {\tt AIPS}.\\
\noindent$^g$ Linear dimensions (major $\times$ minor axis) of the deconvolved emission assuming a distance of 29.9 Mpc \citep{cappellari+11}.\\
\noindent$^h$ Total integrated flux density and error.  The error shown is the sum of the error reported by JMFIT in {\tt AIPS} and the standard VLBA 5\% calibration error, added in quadrature.\\
\noindent$^i$ Log of the radio power assuming a distance of 29.9 Mpc.\\
\noindent$^j$ Brightness temperature, $T_{\rm b} \equiv (S/\Omega_{\rm beam})\frac{c^2}{2k\nu^2}$; where $\nu$ is the observing frequency, $S$ is the integrated flux density, and $\Omega_{\rm beam}$ is the beam solid angle.  The constants $c$ and $k$ are the speed of light and the Boltzmann constant, respectively.\\
\noindent$^k$ Right ascension of the central position of the emission as determined by JMFIT in {\tt AIPS}.  The format is sexagesimal and the epoch is J2000.  The positional uncertainty is $\approx 1$~mas and is dominated by the positional uncertainty of the phase calibrator.\\
\noindent$^l$ Declination of the central position of the emission, determined in the same manner as the right ascension, above.\\
\end{table}

\subsection{VLBA Continuum Source}
\label{results:vlba}
\subsubsection{Missing VLA Flux Density}
\label{missing_flux}
While the VLBA observations may support a scenario in which radio-mode AGN feedback is driving the outflow in NGC~1266, structural information necessary to directly implicate the deposition of jet mechanical energy at the launch-point of the molecular outflow is still lacking.  Additionally, the 1.4~GHz core integrated flux density at arcsecond-scale resolution is nearly two orders of magnitude higher than the level detected by the VLBA.  Compared with the continuum core flux density measured from the line-free channels of our H{\tt I} absorption cube, the VLBA has only recovered about 2\% of the core VLA flux density.  Some of this missing flux density may have an extended, radio outflow origin and could have been resolved-out\footnote{Interferometers act as spatial filters, and are only sensitive to emission over a finite range of spatial scales.  For the VLBA observations of NGC~1266 at 1.65~GHz, the highest possible angular resolution is $\approx$5~mas and is set by the longest baseline length of 8,611~km.  Additionally, the VLBA is blind to structures with spatial extents greater than about 90~mas in our NGC~1266 data.  This limitation on the ``largest angular scale" the emission can have and still be detectable is set by the shortest baseline length, which for the VLBA is 236~km.} by the VLBA, a problem known to plague other low-level AGNs observed at milliarcsecond scales (e.g., NGC~4395; \citealt{wrobel+01, wrobel+06}).  Other possibilities for the origin of the resolved-out emission include shocks or SF in the central molecular disk.


A potential tool to help determine whether SF could account for the missing radio emission is the radio-far-infrared (radio-FIR) relation (e.g., \citealt{helou+85, condon+92, yun+01}).  The radio-FIR relation is believed to arise from recent SF, with thermal infrared dust emission from young stars producing the FIR emission, and nonthermal synchrotron emission associated with supernova remnants (SNRs) producing the low-frequency radio emission.  The radio-FIR relation is defined as follows:

\begin{equation}
\label{eq:q}
q \equiv \log \left( \frac{\mathrm{FIR}}{3.75 \times 10^{12} \, \mathrm{W} \, \mathrm{m}^{-2}} \right) - \log \left(\frac{S_{1.4 \, \mathrm{GHz}}}{\mathrm{W} \, \mathrm{m}^{-2} \, \mathrm{Hz}^{-1}} \right)
\end{equation}

where $\mathrm{FIR} \equiv 1.26 \times 10^{-14} \, (2.58 S_{60 \, \mu m} + S_{100 \, \mu m})$ W~m$^{-2}$ and $S_{60 \, \mu m}$ and $S_{100 \, \mu m}$ are the Infrared Astronomical Satellite (IRAS) 60 and 100~$\mu$m band flux densities in Jy, respectively \citep{yun+01}.  Many studies have analyzed the radio-FIR relation for their respective samples of galaxies and a number of different average q-values for star-forming galaxies have been suggested (see Sargent et al.\ 2010 for a summary).  One of the more widely-cited studies, Yun et al.\ (2001), found an average q-value of 2.34 and defined ``radio-excess'' galaxies (likely AGN-hosts) as those with q $<$ 1.64 and ``FIR-excess'' galaxies (dusty starbursts or dust-enshrouded AGNs) as those with q $>$ 3.00.  For NGC~1266, the NRAO VLA Sky Survey (NVSS; \citealt{condon+98}) 1.4~GHz flux density of 112~mJy and the IRAS flux densities\footnote{The NGC~1266 60 and 100 $\mu$m IRAS flux densities are $S_{60 \, \mu m}$ = 12.80 $\pm$ 0.04 Jy and $S_{100 \, \mu m}$ = 16.90 $\pm$ 0.22 Jy.  Thus, FIR = 6.29 $\times$ 10$^{-13}$ W~m$^{-2}$.} provided in Gil~de~Paz et al.\ (2007) yield q = 2.18.  According to Yun et al.\ (2001), such a q-value would be well within the range of normal star-forming galaxies.

However, several recent studies have shown that the radio-FIR relation can only distinguish normal star-forming galaxies from those which harbor AGNs in the most extreme cases, such as radio-loud quasars \citep{obric+06, mauch+07, moric+10}.  For instance, Obri\'{c} et al.\ (2006) conclude that many confirmed low-luminosity AGNs (LLAGNs; see Section~\ref{accretingbh}), such as Seyferts and LINERs\footnote{A LINER is defined as a ``low-ionization nuclear emission-line region," and most LINERs are believed to harbor LLAGNs.  See Ho 2008 for review.}, have q-values within the scatter of normal star-forming galaxies.  As a quantitative example, Mori\'{c} et al.\ (2010) found an average q-value for all galaxy types in their radio-optical-infrared sample of q = 2.27, with Seyfert galaxies and LINERs yielding average q-values of 2.14 and 2.29, respectively.  Thus, in the context of radio-FIR relation studies which have considered LLAGNs, NGC~1266's q-value is consistent with Seyferts and LINERs.  Other means of assessing the SF contribution to the missing radio emission are therefore necessary.


Although the radio-FIR relation appears to be a rather blunt tool, it has been formulated as an empirical radio SFR indicator (e.g., \citealt{condon+92, yun+01}).  Equation~13 of Yun et al.\ (2001) predicts an SFR of $\approx$6.8 M$_{\odot}$ yr$^{-1}$ for the total 1.4~GHz flux density in NGC~1266 of 112~mJy ($P_{\mathrm{radio}}$ = 1.2 $\times$ 10$^{22}$ W Hz$^{-1}$) from NVSS.  For comparison, the theoretically-derived radio SFR relation provided in Equation~15 of Murphy et al.\ (2011) predicts a slightly higher 1.4~GHz SFR of $\approx$8.5 M$_{\odot}$~yr$^{-1}$.  High-frequency radio continuum emission, which arises from thermal emission from H{\tt II} regions that have been ionized by massive young stars, offers a more direct probe of recent SF.  The 33~GHz flux density of NGC~1266 as measured by the Green Bank Telescope is 8.45 $\pm$ 0.45 mJy \citep{murphy+12}, which corresponds to a radio power of 9.04 $\times$ 10$^{20}$ W Hz$^{-1}$ and an SFR of $\approx$3.7 M$_{\odot}$ yr$^{-1}$ \citep{murphy+11}.  The 33~GHz SFR is consistent with the SFRs derived from other tracers in Alatalo et al.\ (2011), such as the H$_{2}$ surface density (3.1 M$_{\odot}$ yr$^{-1}$) and the FIR luminosity (2.2 M$_{\odot}$ yr$^{-1}$).  However, the VLA 1.4~GHz SFR estimate (6.8$-$8.5 M$_{\odot}$ yr$^{-1}$) is significantly higher.  This enhancement in lower-frequency, non-thermal emission may be due to a contribution from the AGN.  Very long baseline interferometric observations with higher sensitivity and baseline lengths intermediate to those probed by the VLA and VLBA will ultimately be needed to search for the presence of faint, extended, and perhaps jet-like features in the NGC~1266 nucleus.  Our upcoming High Sensitivity Array (HSA) observations may therefore help reveal the origin of the missing VLA flux density.

\begin{figure*}[t!]
\centering
\includegraphics[width=7in]{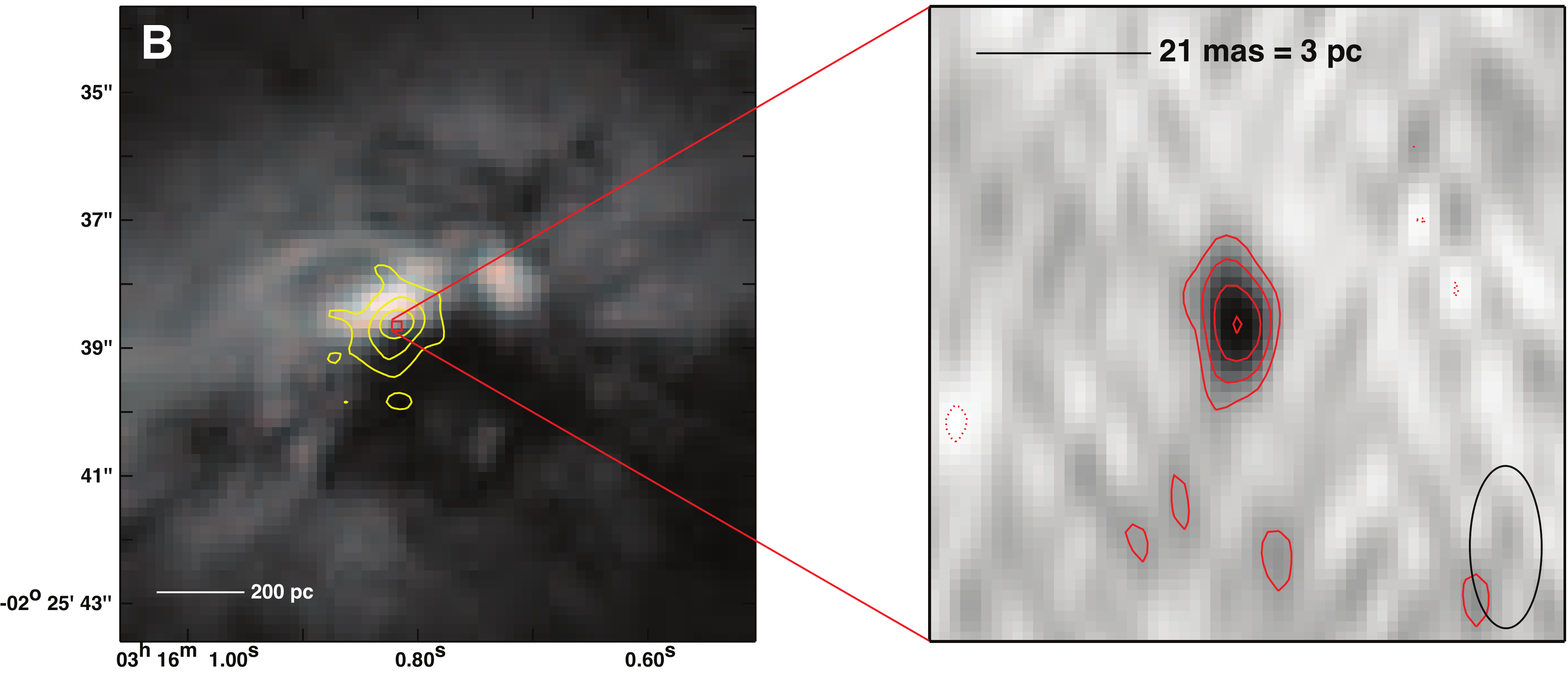}
\caption{{\bf (Left):} A zoomed-in view of the HST {\em B}-band image, plotted to emphasize the enhancement in flux to the north-northwest of the dust cone (Deustua et al., in preparation).  The CARMA CO(1--0) $v_{\rm sys}$ emission presented in Alatalo et al.\ (2011) is overlaid in yellow contours, at levels of \hbox{[0.25, 0.5, 0.75]} of the peak CO emission.  The central box shows the relative field-of-view of the VLBA image.  {\bf (Right):} NGC~1266 1.65~GHz VLBA radio continuum image overlaid with contours.  The VLBA beam is the black ellipse in the lower right corner and has a major axis diameter of 1.14~pc.  The relative contour levels are \hbox{[-3, 3, 6, 10, 14]} and the unit contour level (rms noise) is 42 $\mu$Jy beam$^{-1}$.  The emission is detected to 14$\sigma$ significance, has a brightness temperature above 10$^7$~K, and likely originates from low-level accretion onto a central supermassive black hole.}
\label{fig:vlba}
\end{figure*}

\subsubsection{Compact Starburst}
The VLBA-detected radio emission in the nucleus of NGC~1266 could conceivably originate from a compact starburst, an SNR, or accretion onto the central supermassive black hole (SMBH).  Since compact nuclear starbursts are limited to \hbox{$T_{\rm b}~\lesssim~10^5$}~K \citep{condon+91}, the detection of a high brightness temperature radio core beyond this limit is generally considered strong evidence for accretion onto a black hole.  The VLBA source in NGC~1266 exceeds this limit by over two orders of magnitude, thus excluding the possibility of a compact starburst origin.  

\subsubsection{Supernova Remnant}
Although a current burst of nuclear SF could not generate the observed high brightness temperature emission, synchrotron emission from a single young supernova remnant (SNR) must be considered.  The spectral index of Cas~A of $\alpha$ = $-0.77$ \citep{baars+77} is comparable to the radio spectral index in the nucleus of NGC~1266 of $\alpha = -0.66$ from archival VLA observations (\citealt{davis+12}; Section~\ref{results:vla_archival} of this work).  However, the radio power of Cas~A is only $\sim$10$^{18}$ W~Hz$^{-1}$, two orders of magnitude below the radio power of NGC~1266 measured by the VLBA.  Furthermore, the source in NGC~1266 is considerably more compact (d~$<$~1.2~pc) than Cas~A (d~$\approx$~5~pc).  Due to these disparities in spatial extent and luminosity, a Cas~A-like SNR cannot explain the observed high-resolution radio continuum emission in the nucleus of NGC~1266.

Although a Cas~A analog can be ruled out, a young compact SNR, similar to J1228+441 located in the nearby starburst galaxy NGC~4449, could potentially generate the parsec-scale radio continuum emission in NGC~1266.  J1228+441 is located at a distance of 3.8~Mpc \citep{annibali+08} and is 60$-$200 years old \citep{lacey+07}, with recent analyses suggesting an age closer to 70 years old \citep{bietenholz+10}.  As would be expected for a young SNR, J1228+441 is compact with a reported spatial extent of 1.20~pc $\times$ 0.74~pc \citep{bietenholz+10}, comparable to the upper limit of the extent of the emission in NGC~1266 of 1.16~pc $\times$ 0.90~pc.  Extensive radio observations of J1228+441 have been carried out since the 1970s and have revealed a steady decline in the radio flux density, along with a variable radio spectral index ranging from $-0.6$ to $-1.0$ from 1973 to 2002 \citep{lacey+07}.  The 1.4~GHz radio power of this luminous SNR as measured in 2008 is 1.1~$\times$~10$^{19}$~W~Hz$^{-1}$ \citep{bietenholz+10}, about an order of magnitude weaker than the compact radio source in NGC~1266.  However, if we extrapolate the 2008 1.4~GHz radio power back by 70 years using the fractional decline rate from Lacey et al.\ (2007) ($dS/Sdt = -2.8 \pm 0.4\%$ yr$^{-1}$), we find an estimated radio power of $\approx$8 $\times$ 10$^{19}$ W~Hz$^{-1}$, less than a factor of 2 smaller than the VLBA-detected continuum source in NGC~1266.   

These radio size, spectral index, and luminosity constraints make it difficult to completely rule out a young, compact SNR origin for the nuclear source in NGC~1266 detected with the VLBA.  However, the unusually high luminosity of J1228+441 may be the product of the extreme conditions of its host galaxy, NGC~4449.  The Magellanic-type galaxy NGC~4449 is one of the most luminous and active irregular galaxies known and the only local galaxy with a global starburst \citep{annibali+08}.  NGC~4449 hosts around 60 star clusters, including a young, central super star cluster with a diameter of $\approx$0.1$^{\prime \prime}$, a mass of a few $\times$~10$^{5}$ M$_{\odot}$, and an estimated age of 6$-$10 Myr \citep{boker+01, gelatt+01, annibali+11}.  

NGC~4449 exhibits numerous signs of a recent interaction or merger, including extended gas filaments and streamers, as well as kinematic signatures such as gas counterrotation \citep{annibali+08, annibali+11}.  The young SNR J1228+441 is located within a rich cluster of high-mass OB stars (along with some Wolf Rayet stars) only a few parsecs in size \citep{milisavljevic+08}.  The progenitor star itself is believed to have had a mass $\ga$~20~M$_{\odot}$.  The SNR's extraordinary radio luminosity is believed to be the result of shock interactions with dense circumstellar material from its progenitor star and the winds of massive stellar neighbors.

While NGC~4449 provides a fertile environment for unusually luminous SNRs, the isolated, lenticular, poststarburst galaxy NGC~1266 may not be the most likely host for an SNR similar to J1228+441.  Although the nucleus of NGC~1266 does contain a young stellar population \citep{alatalo+13}, evidence for a dense, super star cluster-like environment or the presence of high-mass stars with masses $\ga$~20~M$_{\odot}$ is currently lacking.  Additionally, the single stellar population age estimate in the central kpc of NGC~1266 is about 1.1~Gyr \citep{alatalo+13}, significantly older than the age of the super star cluster that produced J1228+441.  Since radio continuum emission from young compact SNRs like J1228+441 is known to decline steadily, continued radio monitoring of the NGC~1266 nucleus will help to definitively rule out the possibility of a compact, young SNR. 

\subsubsection{Accreting Supermassive Black Hole}
\label{accretingbh}
The final possibility for the origin of the VLBA continuum source is emission from accretion onto the central SMBH.  Given the abundance of gas in the vicinity of the nucleus of NGC~1266 that could potentially fuel the SMBH, the presence of a low-level AGN is not unexpected.  From the central stellar velocity dispersion of $\sigma \approx$ 79 km~s$^{-1}$ \citep{cappellari+13} and the $M_{\mathrm{BH}}-\sigma$ relation \citep{gultekin+09}, we estimate that the SMBH at the center of NGC~1266 has a mass of $\approx$2.57 $\times$ 10$^{6}$ M$_{\odot}$ (which ranges from 9.33 $\times$ 10$^{5}$ to 7.08 $\times$ 10$^{6}$ M$_{\odot}$ when the intrinsic scatter of 0.44 is taken into account).  Thus, the relatively modest mass estimate of the central SMBH in NGC~1266 is similar to the mass of Sgr A$^*$ in the Galactic Center of 4.31 $\pm$ 0.36 $\times$ 10$^6$ M$_{\odot}$ \citep{gillessen+09}.

Although NGC~1266 and Sgr A$^*$ may harbor SMBHs with similar masses, NGC~1266 is a significantly more luminous radio source.  The VLA 5~GHz emission of NGC~1266 (Table~\ref{tab:vla_archival}) has a nuclear radio power of 2.17~$\times$~10$^{21}$~W~Hz$^{-1}$, nearly 6 orders of magnitude higher than the average 5~GHz radio power of Sgr A$^*$ ($\approx$4.59 $\times$ 10$^{15}$ W Hz$^{-1}$; \citealt{melia+01, falcke+04}).  The radio power of NGC~1266, while significantly greater than that of Sgr A$^*$, is more similar to nearby LLAGNs, which typically have radio powers in the range 10$^{18}-10^{25}$ W~Hz$^{-1}$ \citep{ho+01, ho+08}, than to the powerful quasar hosts typically associated with massive outflows both observationally (e.g., \citealt{cano-diaz+12}) and in simulations (e.g., \citealt{debuhr+12}).  Although Ho (2002) failed to find a relationship between 5~GHz radio power and black hole mass, the average 5~GHz radio power of the 7 Seyfert nuclei with black hole masses in the range expected for NGC~1266 is about 4 $\times$ 10$^{21}$ W~Hz$^{-1}$, similar to the 5~GHz radio power of NGC~1266.

The archival VLA radio spectral index measurements (Section~\ref{results:vla_archival}) are comparable to those of other LLAGNs, which can range from $\alpha$ = 0.5 to $-1.0$ between 1.4 and 5 GHz (e.g., \citealt{ho+01}).  The VLBA detection at 1.65~GHz implies an even more modest radio power of 1.48~$\times$~10$^{20}$~W~Hz$^{-1}$, corresponding to $\log T_{\mathrm{b}} \gtrsim$ 7.18.  This brightness temperature is consistent with other VLBA surveys of LLAGNS, which have found brightness temperature lower limits of $\log T_{\mathrm{b}} \gtrsim$ 6.3 to $\log T_{\mathrm{b}} \gtrsim$ 10.8. (\citealt{nagar+05}).

Although direct observational evidence that LLAGNs may be capable of shutting down late-time SF in the local universe is still lacking, recent empirical arguments have supported this scenario.  For instance, Schawinski et al.\ (2007) performed an analysis of 16,000 early-type galaxies from the Sloan Digital Sky Survey (SDSS; \citealt{york+00}) and found that many local early-type galaxies show a clear time sequence as they transition from actively star-forming to a composite phase (in which SF and an AGN coexist) and finally to an AGN-dominated phase.  Schawinski et al.\ (2009) performed a follow-up analysis of the cold gas content of local early-type galaxies and found that molecular gas disappeared within 100 Myr of the onset of a low-level AGN episode.  Kaviraj et al.\ (2011) further supported these empirical findings through their phenomenological model.  This model argued that the depletion of cold gas by SF alone is too slow to account for the rapid transition from the blue cloud to the red sequence inferred from SDSS studies.  Thus, the quenching of SF by an LLAGN in NGC~1266 would confirm that even modest nuclear activity can be an important driver of galaxy evolution at the current epoch.

\section{Discussion}
\label{disc}
Alatalo et al.\ (2011) argued that the SFR in the nucleus of NGC~1266 is likely insufficient to drive the observed molecular outflow, and cited AGN feedback as the most probable driving mechanism.  Although the VLBA continuum source could conceivably originate from a young SNR or an LLAGN, the high brightness temperature eliminates the compact starburst scenario.  A young, exceptionally luminous SNR, though it cannot be ruled out completely, is unlikely given the extreme environment required for such an SNR.  Therefore, the VLBA continuum emission most likely originates from the central low-level AGN.  In the following section, we thus re-examine the ability of the AGN to drive the outflow of cold gas.

\subsection{Radio Continuum Energetics}
\subsubsection{Minimum Jet Energy}
\label{energy}
Although the detailed physics of AGN-driven outflows are still poorly understood (e.g., \citealt{fabian+12}), a number of studies have attempted to quantify radio jet energy and power, both theoretically and observationally.  Since we expect nonthermal radio emission to arise from synchrotron emission, we can use our knowledge of the underlying physics to estimate the minimum energy available in the radio jet to drive the outflow based on the measured radio luminosity, spectral index, and spatial extent (e.g., \citealt{begelman+84, miley+80, moffet+75}).  Following the derivation in Moffet et al.\ (1975), the total energy arising from synchrotron emission in a radio source is:

\begin{equation}
\label{eq:totalenergy}
U_\mathrm{T} = U_\mathrm{p} + U_\mathrm{m} = aALB^{-3/2} + VB^2/8\pi
\end{equation}

\noindent where $U_T$ is the total energy of the source, $U_\mathrm{p}$ is the total particle energy, and $U_\mathrm{m}$ is the total magnetic field energy.  $U_\mathrm{p} \equiv aU_\mathrm{e}$, where $a$ is the contribution of protons relative to electrons to the total particle energy and $U_\mathrm{e}$ is the total energy in the relativistic electrons.  $L$ is the luminosity of the source, $B$ is the average magnetic field strength, and $V$ is the volume of the synchrotron emitting region.  The term $A = C \frac{2\alpha + 2}{2\alpha + 1} \frac{\nu_{2}^{\alpha + 1/2} - \nu_{1}^{\alpha + 1/2}}{\nu_{2}^{\alpha + 1} - \nu_{1}^{\alpha + 1}}$, where $C$ is a constant\footnote{the cgs units of $C$ are (g/cm)$^{3/4}$ s$^{-1}$.} of value 1.057 $\times$ 10$^{12}$, $\alpha$ is the spectral index, and $\nu_1$ and $\nu_2$ are the assumed lower and upper frequencies of the radio spectrum, respectively.  Equation~\ref{eq:totalenergy} is minimized near the ``equipartition" value of the magnetic field, where $U_\mathrm{p}$ and $U_\mathrm{m}$ are nearly equal, leading to $U_{\mathrm{min}}$ = 0.5$(aAL)^{4/7}V^{3/7}$ and $B_{\mathrm{min}}=2.3(aAL/V)^{2/7}$.  

For NGC~1266, we assume $a$ = 100 \citep{moffet+75}, $\nu_1$ = 10~MHz, $\nu_2$ = 100~GHz, $L$ = 3.0 $\times$ 10$^{37}$ erg s$^{-1}$ (the total 1.4~GHz luminosity in the lobes, measured by summing the flux in each lobe using the Viewer tool in version 4.0.0 of the Common Astronomy Software Applications package), $\alpha$ = $-$0.79, and lobe radii of 0.12 and 0.34~kpc for the northern and southern lobes, respectively.  This provides an estimated minimum energy in the radio lobes of $U_{\mathrm{min}}$ = 1.7 $\times$ 10$^{54}$ erg, corresponding to an equipartition magnetic field of $B_{\mathrm{min}}$ = 50 $\mu$Gauss\footnote{The equipartition magnetic field of NGC~1266 is within the range of typical estimates (1 to 100 $\mu$Gauss; \citealt{kellermann+88}).}.

We can compare this minimum jet energy to the kinetic energy of the outflow.  Assuming a cold gas outflow mass of 3.3 $\times$ 10$^7$ M$_{\odot}$ (which includes H$_2$ and H{\tt I}), the kinetic energy of the outflow is about 1.0 $\times$ 10$^{55}$ erg (\citealt{alatalo+11}).  The difference between the energy in the outflow and the minimum energy of the lobes is about an order of magnitude, similar to that observed in M51 \citep{matsushita+04}.  Therefore, the true energy of the lobes, which may be significantly higher than our minimum energy estimate, may be sufficient to entrain the molecular gas and drive it out of NGC~1266.  Future studies of the radio spectrum and measurements of the magnetic field strength and geometry in NGC~1266 will allow for a more accurate estimation of the jet energy \citep{hardcastle+04}.

\subsubsection{Jet Mechanical Power}
\label{power}
Interestingly, many theoretical models aimed at describing the compact flat-spectrum radio emission produced at the launch point of a relativistic jet actually predict that radio power ($P_{\mathrm{radio}}$) and jet mechanical power ($P_{\mathrm{jet}}$) should be correlated (e.g., \citealt{blandford+79, heinz+03}).  Samples of AGNs with X-ray cavities presumably inflated by radio jets can be used to calibrate the $P_{\mathrm{radio}}-P_{\mathrm{jet}}$ relation \citep{birzan+04, birzan+08, cavagnolo+10} so that it can be used to predict the mechanical powers of jets based on radio continuum information alone\footnote{We caution that empirically-derived $P_{\mathrm{jet}}$ estimates have large intrinsic uncertainties due to factors such as unknown radio source ages, differences in radiative efficiencies, varying magnetic field strengths, and the heterogeneous radio morphologies of sample sources.  See B\^{i}rzan et al.\ 2008 for review.}.

Merloni \& Heinz (2007) studied the statistical relationship between core $P_{\mathrm{radio}}$ at 5~GHz and $P_{\mathrm{jet}}$ for a sample of sources with measured jet mechanical powers and found\footnote{Note that in this relation (Equation~6 in Merloni \& Heinz 2007), the intrinsic radio luminosity, $L_{\mathrm{radio}}$, has been estimated using a statistical correction for the effects of relativistic beaming in the sample.} $P_{\mathrm{jet}} \approx 1.6 \times 10^{36}$ erg s$^{-1}$ $(L_{\mathrm{radio}})^{0.81}$, where $L_{\mathrm{radio}}$ is in units of 10$^{30}$ erg s$^{-1}$ and the intrinsic scatter is $\sigma_{\mathrm{int}}$ = 0.37~dex.  Since the synchrotron emission originating at the launch point of the radio jet is assumed to have a flat radio spectral index, we can use the Merloni \& Heinz (2007) relation to predict the jet mechanical power of NGC~1266 based on our 1.65~GHz VLBA detection.  With a radio power of 2.44 $\times$ 10$^{36}$ erg~s$^{-1}$, the VLBA emission implies a jet mechanical power of $\approx$2.4 $\times$ 10$^{41}$ erg~s$^{-1}$ (1.0 $-$ 5.5 $\times$ 10$^{41} $ erg~s$^{-1}$).  The mechanical luminosity of the outflow is $L_{\mathrm{mech}} \approx 1.3 \times 10^{41}$ erg s$^{-1}$ \citep{alatalo+11}, less than our estimate of the jet mechanical power implied by the VLBA detection.  

This suggests that the AGN in NGC~1266 may be powerful enough to drive the outflow, albeit with a relatively large coupling of $\approx$54\% (24\% to more than 100\%) between the total mechanical energy of the jet and the energy transfered to the ISM.  Although this coupling fraction is roughly consistent with predictions from recent simulations of jet feedback by Wagner et al.\ (2012) (10$-$40\%), the jet mechanical power in NGC~1266 is significantly lower than the range of $P_{\mathrm{jet}}$ expected to produce effective feedback (10$^{43}-10^{46}$ erg~s$^{-1}$; \citealt{wagner+12}).  Thus, the ability of a putative AGN jet in NGC~1266 to drive the observed outflow remains a possibility, but subsequent observations aiming to more deeply explore the nuclear ISM conditions and resolve the compact radio emission will be needed to fully settle this issue.

For comparison, the $P_{\mathrm{radio}}-P_{\mathrm{jet}}$ relation can also be derived directly from samples of sources with X-ray cavities carved out by radio jets.  The studies by B\^{i}rzan et al.\ (2004, 2008) were the first to calculate such a relation, in which they utilized the total 1.4~GHz continuum emission (core + jets) at much lower spatial resolution than the Merloni \& Heinz (2007) study.  Since the sample used by B\^{i}rzan et al.\ (2004, 2008)  was dominated by galaxy clusters with high radio luminosities, we use the Cavagnolo et al.\ (2010) relation ($P_{\mathrm{jet}} \approx 5.8 \times 10^{43}$ erg~s$^{-1}$ $(L_{\mathrm{radio}})^{0.7}$, where $L_{\mathrm{radio}}$ is in units of 10$^{40}$ erg~s$^{-1}$ and the intrinsic scatter is $\sigma_{\mathrm{int}}$ $\approx$ 0.7~dex) instead, which was derived in the same basic way but included a more balanced sample of galaxy clusters and lower radio power sources (e.g., elliptical galaxies).

In order to be consistent with the large spatial scales of the radio continuum emission considered in the Cavagnolo et al.\ (2010) study, we used the NVSS integrated flux density of 112~mJy.  Thus, for the total NVSS radio power of 1.7 $\times$ 10$^{38}$ erg~s$^{-1}$, the Cavagnolo et al.\ (2010) relation predicts a jet mechanical power of $\approx$3.3 $\times$ 10$^{42}$ erg~s$^{-1}$ (6.6 $\times$ 10$^{41}$ $-$ 1.7 $\times$ 10$^{43} $ erg~s$^{-1}$), implying a jet-ISM coupling of $\approx$4\% (0.8$-$20\%).  This estimate is significantly larger than the $P_{\mathrm{jet}}$ estimate from {\color{blue} Merloni \& Heinz (2007)}, suggesting a smaller jet-ISM coupling, and is more in line with the range of jet powers predicted to effectively induce feedback in {\color{blue} Wagner et al.\ (2012)}. 

The results of our jet power estimates based on the compact VLBA emission and total NVSS flux density further support the scenario in which feedback from the AGN is driving the outflow in NGC~1266, although future radio continuum and ISM studies will be necessary to improve current jet power estimates.  Our new jet power estimates update the one provided in Section~4.3 of Alatalo et al.\ (2011).

\subsection{Does NGC~1266 Harbor a Compton-thick AGN?}
Figures~\ref{fig:co+dust} and \ref{fig:vlba} show that the VLBA continuum source lies within the densest portion of the CO emission. The detailed analysis of the CO emission presented in Alatalo et al.\ (2011) indicated an H$_2$ column density of $1.7\times10^{24}$ cm$^{-2}$ in the nucleus of NGC~1266.  This thick column of molecular gas appears to be sitting along the line-of-sight to the central AGN, likely heavily obscuring the emission.  Subsequent studies have revealed even higher molecular gas column densities (\citealt{crocker+12, bayet+13}; Alatalo et al., in preparation).

{\em Chandra} observations of NGC~1266 from 2010 revealed diffuse and mainly soft X-ray emission, likely the result of thermal Bremsstrahlung in regions of shocked ISM due to the outflow \citep{alatalo+11}.  Although X-ray cores are commonplace in LLAGNs \citep{ho+08}, NGC~1266 does not appear to host compact, nuclear, hard X-ray emission.  A weak detection of hard X-rays in the 4-8 keV regime near the nucleus and coincident with the archival VLA 5~GHz emission was reported in Alatalo et al.\ (2011), but this emission was not point-source-like and therefore may not originate from accretion onto the SMBH.  This observed dearth of compact, hard nuclear X-ray emission may be the result of the high column density.  In fact, the H$_2$ column density alone exceeds the formal limit for a Compton-thick absorber of $N_{\mathrm{H}}~\ga~1.5~\times~10^{24}$ cm$^{-2}$ \citep{comastri+04}, implying that the AGN in NGC~1266 may be mildly Compton-thick.  While a more detailed analysis of the existing NGC~1266 {\em Chandra} data will be presented in a follow-up paper, more sensitive X-ray studies will be needed to better measure the X-ray spectrum of the AGN buried within NGC~1266, and verify that it is indeed Compton-thick.

If NGC~1266 harbors a bona fide Compton-thick AGN, it will join the few dozen known local AGNs in this category \citep{comastri+04}.  Despite the paucity of local examples, recent surveys have suggested that a significant fraction of AGNs are in fact Compton-thick \citep{risaliti+99, malizia+09, treister+09}.  For instance, Malizia et al.\ (2009) reported that out of their sample of 25 AGNs within 60~Mpc detected by the International Gamma-Ray Astrophysics Laboratory (INTEGRAL), 24\% were Compton-thick.  

Thus, other galaxies with large column densities similar to NGC~1266 may be missed in AGN surveys that rely on the detection of X-ray emission.   In the case of NGC~1266, the VLBA continuum detection was essential in supporting the presence of the AGN, and we suggest that radio continuum searches should play a prominent role in the detection and study of obscured AGNs.\\

\section{Summary and Conclusions}
\label{conclu}
We report new VLA H{\tt I} absorption and VLBA continuum observations of the AGN-driven molecular outflow candidate NGC~1266.  The high spatial resolution H{\tt I} absorption is compact and associated with the core 1.4~GHz continuum emission, but not the extended radio lobes.  The observed compactness of the H{\tt I} absorption could be due to the fact that the H{\tt I} phase of the outflow is truly localized near the central SMBH or due to sensitivity limitations.  The H{\tt I} absorption spectrum is best fit by a double Gaussian, with a deep component near the systemic velocity and a shallow blueshifted component associated with the outflow with a mass of 1.4 $\times$ 10$^{6}$ M$_{\odot}$ and a relative velocity of $-141.7$ km s$^{-1}$.  Based on the compact morphology of the H{\tt I} absorption and its relatively low outflow velocity, it is currently not clear whether the radio jet is driving this phase of the outflow.  Future studies of H{\tt I} absorption in lower radio power sources will be necessary to determine whether jets produced by local, lower-luminosity AGNs can effectively impact their host galaxies.

The VLBA detection of compact emission further strengthens the case for the presence of an AGN in NGC~1266.  The VLBA-derived brightness temperature of $T_{\mathrm{b}}~\ga~1.5~\times~10^7$~K at 1.65~GHz is far too high to be explained by a compact nuclear starburst.  Although an unusually luminous, young, compact SNR in a dense nuclear super star cluster with a high-mass progenitor star could conceivably produce the observed VLBA emission, we consider this scenario unlikely given the extreme conditions necessary to produce such an SNR.  Thus, the most likely scenario is that the VLBA emission originates from accretion onto an SMBH.

The parsec-scale radio continuum emission has a power of 1.48 $\times$ 10$^{20}$ W~Hz$^{-1}$, on par with the typical radio powers of LLAGNs on milliarcsecond scales.  However, compared with lower-spatial-resolution VLA data at 1.4~GHz, only $\approx$2\% of the total low-frequency continuum emission was detected by the VLBA.  Upcoming interferometric observations with the HSA (consisting of the VLBA and the phased VLA) will provide images with improved sensitivity to faint, extended structures, and may recover some of the ``missing" flux density.

The location of the AGN derived from the VLBA data coincides with the kinematic center of the galaxy and lies within the densest portion of the molecular gas, suggesting that the AGN within NGC~1266 is heavily obscured and may be mildly Compton-thick.  The data presented in this paper have highlighted the role of radio continuum observations in identifying AGNs in heavily obscured environments.  

Jet energy and mechanical power estimates suggest that the AGN in NGC~1266 may indeed be powerful enough to drive the outflow.  The minimum energy of the jet assuming the equipartition value of the magnetic field is about an order of magnitude less than the kinetic energy of the outflow, similar to the situation in other nearby galaxies hosting candidate AGN-driven outflows (e.g., \citealt{matsushita+04}).  Future studies of the radio spectrum as well as magnetic field measurements from radio polarization and deep X-ray observations may help improve the energy estimate of the radio jet in NGC~1266 \citep{hardcastle+04}.  Two independent estimates of the jet mechanical power from empirical relations actually indicate that it exceeds the mechanical luminosity of the outflow.  Thus, this line of evidence does support a scenario in which the AGN is energetic enough to drive the outflow.  

In order to better understand the physical conditions of its obscured AGN, future observational studies of NGC~1266 should focus on resolving the emission detected by the VLBA, determining the radio spectrum, measuring the magnetic field, modeling the X-ray spectrum, and improving existing molecular gas observations.  Improvements to numerical simulations may also help explain how NGC~1266 arrived at its current, unique evolutionary stage.  Continued, deep radio continuum studies, as well as observations with ALMA, will increase the population of known, local AGN-driven outflow candidates.  These studies will help clarify the impact of AGN feedback on star formation in the local universe and will ultimately improve our understanding of galaxy evolution.

\acknowledgments{{\it Acknowledgments:}  We thank the anonymous referee for thoughtful comments that have improved the strength of this work.  The National Radio Astronomy Observatory is a facility of the National Science Foundation operated under cooperative agreement by Associated Universities, Inc.  The Space Telescope Science Institute is operated by the Association of Universities for Research in Astronomy, Inc., under NASA contract NAS5-26555.  Support for CARMA construction was derived from the Gordon and Betty Moore Foundation, the Kenneth T. and Eileen L. Norris Foundation, the James S. McDonnell Foundation, the Associates of the California Institute of Technology, the University of Chicago, the states of California, Illinois, and Maryland, and the National Science Foundation. Ongoing CARMA development and operations are supported by the National Science Foundation under a cooperative agreement, and by the CARMA partner universities.  This research was funded in part by NSF grant 1109803.

{\it Facilities:} \facility{NRAO}, \facility{HST}.
\\

\bibliographystyle{apj}


 \end{document}